\documentclass[%
  reprint,
  superscriptaddress,
  amsmath, amssymb,
  aps, pra,
  floatfix,
]{revtex4-2}

\usepackage[T1]{fontenc}
\usepackage{bm}             % Bold math

\usepackage{float}          % [H] placement unnecessary with revtex
\usepackage{dcolumn}        % Align table columns on decimal point
\usepackage{booktabs}

\usepackage{multirow}

\usepackage{braket}         % Dirac notation

\usepackage{comment}
\usepackage[normalem]{ulem} % \sout etc. without breaking \emph

\usepackage{xcolor}

\usepackage{graphicx}

\usepackage{hyperref}
\hypersetup{
  colorlinks = true,
  linkcolor  = blue,
  citecolor  = blue,
  urlcolor   = blue,
}

\preprint{APS/123-QED}

\begin{document}
\title{Observation of magnetic quantum phase crossovers in a semiconductor spin ladder}
\renewcommand{\thefootnote}{\fnsymbol{footnote}}

\author{Elizaveta Morozova}
\thanks{These authors contributed equally.}
\affiliation{
QuTech and Kavli Institute of Nanoscience, Delft University of Technology, 2600 GA Delft, The Netherlands}

\author{Xin Zhang}
\thanks{These authors contributed equally.}
\affiliation{
QuTech and Kavli Institute of Nanoscience, Delft University of Technology, 2600 GA Delft, The Netherlands}

\author{Utso Bhattacharya}
\thanks{These authors contributed equally.}
\affiliation{
Institute for Theoretical Physics, Wolfgang-Pauli-Strasse 27, ETH Zurich, 8093 Zurich, Switzerland
}
\affiliation{IBM Quantum, IBM Research – Zurich, 8803 R{\"u}schlikon, Switzerland}

\author{Pablo Cova Fari\~{n}a}
\affiliation{
QuTech and Kavli Institute of Nanoscience, Delft University of Technology, 2600 GA Delft, The Netherlands}

\author{Daniel Jirovec}
\affiliation{
QuTech and Kavli Institute of Nanoscience, Delft University of Technology, 2600 GA Delft, The Netherlands}

\author{Alexander Nico-Katz}
\affiliation{
Department of Physics and Astronomy, University College London, London WC1E 6BT, United Kingdoms}

\author{Stefan D. Oosterhout}
\affiliation{
Netherlands Organisation for Applied Scientific Research (TNO), 2628 CK Delft, The Netherlands}

\author{Sougato Bose}
\affiliation{
Department of Physics and Astronomy, University College London, London WC1E 6BT, United Kingdoms}

\author{Giordano Scappucci}
\affiliation{
QuTech and Kavli Institute of Nanoscience, Delft University of Technology, 2600 GA Delft, The Netherlands}

\author{Menno Veldhorst}
\affiliation{
QuTech and Kavli Institute of Nanoscience, Delft University of Technology, 2600 GA Delft, The Netherlands}

\author{Eugene Demler}
\email{demlere@phys.ethz.ch}
\affiliation{
Institute for Theoretical Physics, Wolfgang-Pauli-Strasse 27, ETH Zurich, 8093 Zurich, Switzerland
}

\author{Lieven M. K. Vandersypen}
\email{l.m.k.vandersypen@tudelft.nl}
\affiliation{
QuTech and Kavli Institute of Nanoscience, Delft University of Technology, 2600 GA Delft, The Netherlands}

\date{\today}

\begin{abstract}
Understanding collective phases of strongly correlated quantum magnets relies on theoretically tractable model systems with precise microscopic control. Antiferromagnetic spin ladders provide such a setting, hosting field-tunable gapped and gapless phases at half filling and unconventional pairing tendencies upon doping. Here, we realize a programmable Heisenberg spin ladder in a half-filled germanium quantum dot array featuring site-resolved, continuously tunable exchange interactions. Under a fixed magnetic field, we vary the rung and leg coupling to map the rung-singlet, canted antiferromagnetic, and fully polarized phases. Hamiltonian-learning protocols combining equilibrium and dynamical measurements quantitatively characterize the ladder, incorporating spin-orbit interactions to reproduce the observed crossover behavior. Measurements of higher-order spin correlators—including four-point correlations inaccessible to conventional bulk probes—reveal signatures of the underlying phase structure despite the finite size. Our results establish germanium quantum dot arrays as a controllable platform for quantum magnetism, opening routes to investigate unconventional superconductivity in doped ladders.
\end{abstract}

\maketitle

\section{Introduction}
Antiferromagnetic Heisenberg ladders provide some of the most popular theoretical models of quantum magnetism. They feature rich phase diagrams with quantum phase transitions (QPT) between the gapped and gapless phases that can be controlled using both exchange interactions and magnetic field~\cite{Dagotto1996, OrignacGiamarchi1999}. Gapped phases correspond to quantum disordered paramagnetic states, while gapless phases are closely related to states with spontaneous symmetry breaking in higher dimensions. Yet antiferromagnetic ladders are simple enough to provide accurate theoretical descriptions of phases and quantum phase transitions, using advanced analytical and numerical techniques, including Abelian~\cite{Giamarchi2003} and non-Abelian bosonization~\cite{Gogolin1998} and density-matrix renormalization group methods~\cite{Noack1994}. Thus, transitions between the gapped and gapless phases in ladders are commonly used as textbook examples of QPT, in which one can understand the evolution of correlations and collective excitations. Another important aspect of antiferromagnetic ladders is that upon doping they are expected to give rise to superconductivity, with the character of pairing reminiscent of d-wave pairing in two dimensional systems~\cite{Dagotto1992, BalentsFisher1996, Dagotto1999}. Thus, doped antiferromagnetic ladders have been studied theoretically with the goal of understanding superconductivity in doped two dimensional Mott insulators, such as high-$T_c$ cuprates~\cite{Chudzinski2007,Chudzinski2008, PhysicsofhighTcSupercond}. 
Earlier experimental studies of antiferromagnetic spin ladders have relied on neutron scattering~\cite{Eccleston1996, Notbohm2007} and Raman~\cite{Gozar2001} spectroscopy in bulk samples. While the observed properties are in general agreement with theoretical expectations, many important aspects have not yet been studied experimentally. Bulk systems do not allow continuous tuning of interactions, there are unresolved issues about the role of disorder in real materials~\cite{Azuma1998, Orignac1998, OrignacGiamarchi1999}, and traditional bulk experimental techniques measure only  two-point correlation functions, such as dynamical spin correlation functions measured in neutron scattering  experiments. Some types of collective excitations can not be seen in dynamical spin correlation functions and require non-linear spectroscopy at non-zero momentum ~\cite{DeSantis2026}, which is challenging with traditional probes of bulk materials.

The simplest realization of these ideas is the spin-$1/2$ Heisenberg ladder in a magnetic field, whose ground-state phase diagram has become a textbook example of quantum criticality~\cite{Dagotto1996, Giamarchi2008BECMagneticInsulators, proukakis2017universal, Chou2025SpinLadderQuantumSimulators,Zapf2014BECQuantumMagnets}. In the thermodynamic limit, the model exhibits three distinct ground-state regimes as a function of magnetic field~\cite{CoupledLaddersMagneticField}:
(1) a total-spin-zero gapped phase for $H < H_{C1}$; (2) a gapless phase for $H_{C1} < H < H_{C2}$; and (3) a fully spin-polarized gapped phase for $H > H_{C2}$. The transitions between these phases constitute quantum phase transitions, while the gapless phase, characterized by critical correlations and emergent quasiparticles, has been associated with Tomonaga--Luttinger-liquid behavior and triplon condensation~\cite{Giamarchi2008BECMagneticInsulators}.

In quantum simulation experiments, systems are finite and often include weak symmetry-breaking terms. As a result, sharp phase transitions are replaced by crossovers~\cite{Crossovers2024}. Nonetheless, analyzing the equilibrium properties and excitation spectra in these crossover regimes provides valuable insight into the fundamental behavior of paradigmatic quantum many-body models~\cite{Georgescu2014}.

\begin{figure}[b]
    \centering
    \includegraphics[width=\columnwidth]{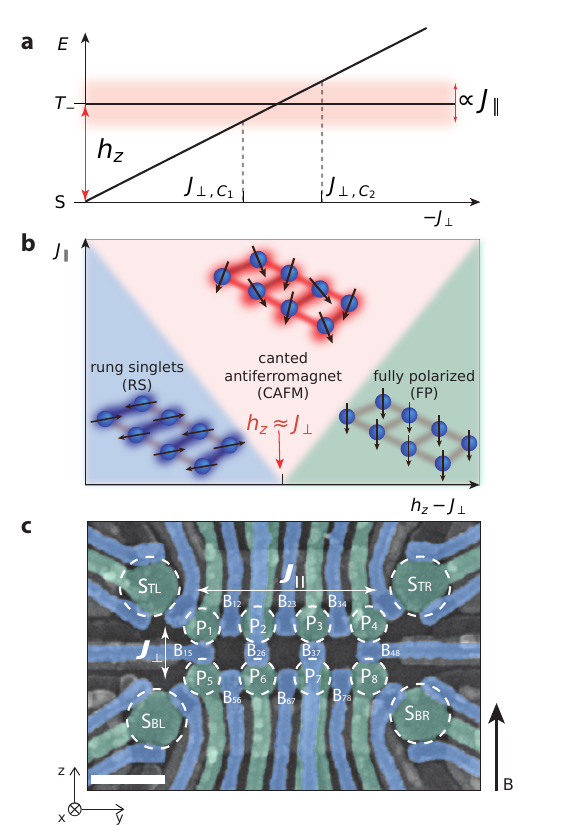}
    \caption{%
    \textbf{Device and experimental concept.}
    (a,b) Schematic illustration of the experiment, showing (a) evolution of energy levels as a function of the exchange coupling along the rungs $J_\perp$ and (b) the three phases of a Heisenberg exchange spin ladder in the thermodynamic limit. Cartoons of a $2\times 4$ ladder illustrate the respective phases. Panel (a) particularly visualizes the relationship of the two critical points in the rung coupling, $J_{\perp, C_1}$ and $J_{\perp, C_2}$, to the leg exchange coupling $J_{\parallel}$.
    (c) False-colored scanning electron microscope image of the device. The device comprises eight sites, defined by plunger gates $P_i$ with $i = 1,\dots,8$, and inter-site barrier gates $B_{ij}$, which tune the exchange coupling between neighboring sites $i$ and $j$. The arrows labeled $J_\perp$ and $J_{\parallel}$ indicate the directions of the exchange interactions. The scalebar is 200 nm. External magnetic field is pointing long $z$-axis.
    }
    % \vspace{-6pt}  
    \label{fig:fig1}
\end{figure}

Semiconductor quantum dots provide a highly tunable platform for quantum  simulation, in which spin-spin interactions can be engineered and coherently controlled at the single-site level~\cite{Hensgens2017QuantumSimulationFermiHubbardQDs,Dehollain2020NagaokaFerromagnetismQD,Kiczynski2022EngineeringTopologicalStates,VanDiepen2021QuantumSimulationHeisenbergChain,Wang2023ProbingRVBGeSimulator,Hsiao2024ExcitonTransportGeLadder,2025ManyBodyInterferometry, hendrickx2021four, SweetSpotGeHoleSpinQubit,seidler2025spatial,wang2024operating, farina2025site, ParraRodriguez2020}. Electrostatic gates enable precise tuning of exchange couplings between neighboring spins, allowing the realization of programmable spin Hamiltonians~\cite{Loss1998QuantumComputationQuantumDots, Hanson2007SpinsFewElectronQuantumDots} beyond static material constraints. The relative strength of exchange and Zeeman energies can be tuned over a broad range~\cite{farina2025site,Zhang2025UniversalControlST,saez2025exchange,farina2025site}, allowing to study the quantum phase diagram across the three expected phases. Among the various semiconductor quantum dot platforms, holes in Ge/SiGe are particularly attractive for forming large quantum dot arrays\cite{Hsiao2024ExcitonTransportGeLadder, farina2025site, 2025ManyBodyInterferometry}, due to their low effective mass, leading to comparatively large quantum dot dimensions and relaxed nanofabrication constraints ~\cite{Scappucci2021GermaniumQuantumInformationRoute}. 
However, depending on the type of physics and geometry to be simulated,  several challenges may arise. For example, the individual calibration, control, and mitigation of crosstalk between Hamiltonian parameters are nontrivial~\cite{Jirovec2025MitigationExchangeCrosstalk}. In addition, high-quality readout of the states of multiple spins (or spin pairs) across an array in every shot still needs to be demonstrated to enable complete characterization of correlated quantum states. Finally, the impact of the strong spin-orbit interaction (SOI) for holes in Ge/SiGe and the resulting variability of the spin quantization axis on the respective phases and the transitions between them needs to be clarified.

Here we demonstrate that the essential characteristics of the quantum phases in a Heisenberg spin ladder can be accessed in a small-scale semiconductor quantum dot array. We first introduce the theory of phase crossovers in a quantum dot ladder in the presence of SOI. We then present experimental results that examine the theoretical predictions, revealing the continuous evolution of the ground state across three distinct magnetic phases as a function of both rung and leg couplings. We further probe four-point correlators, which provide information on the nature of the three phases beyond the two-point correlators traditionally accessible in neutron-scattering experiments~\cite{Eccleston1996, Notbohm2007}.

\section{Quantum Phase Crossover: Theory}

\begin{figure*}[t]
    \centering
    \includegraphics[width=1.0\textwidth]{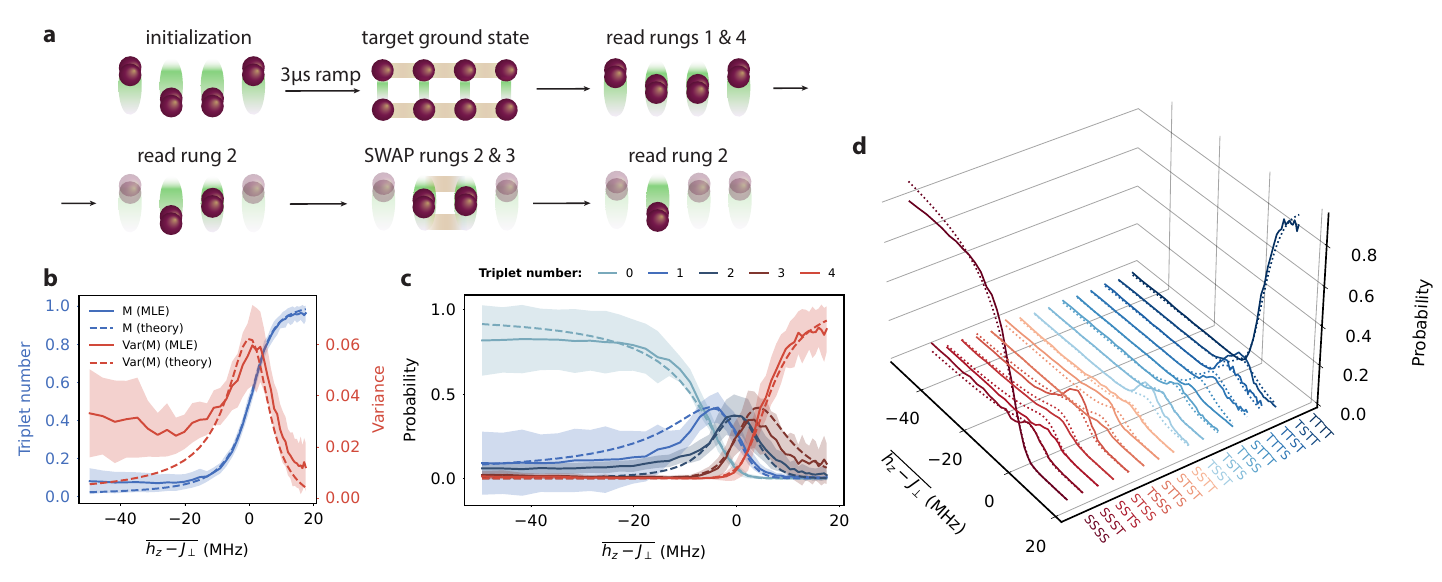}
    \caption{%
    \textbf{Quantum phase crossover in the limit of weak leg coupling.}
    (a) Pulse sequence used in the experiment (Supplementary
Section ~\ref{sec:pulse}).
    (b) Evolution of the triplet number and its variance as a function of $\overline{(h_z - J_{\perp})}$ at $J_{||}  \approx 1 $ ~MHz. 
    The shaded area indicates the standard deviation across the 60 experimental repetitions. Dashed lines show the theoretically predicted values obtained by solving the Hamiltonian in Eq.~\ref{Eq:mainH1}.
    (c) The same data set analyzed in terms of the triplet-number distribution as a function of $\overline{(h_z - J_{\perp})}$. Dashed lines show the theoretically predicted distributions obtained by solving the Hamiltonian in Supplementary Section~\ref{sec:hamiltonian},  Eq.~\ref{Eq:mainH1}.
    (d) Full probability distribution over all rung configurations as a function of the spin-gap parameter. Dashed lines show the theoretically predicted probabilities based on Eq.~\ref{Eq:mainH1}.
    In panels (b-d), 300 single-shot measurements are acquired for each value of spin-gap parameter $\overline{(h_z - J_{\perp})}$, and the full experiment is repeated 60 times.
    }
    \label{fig:fig2}
\end{figure*}

In the low-energy subspace of spin ladder with $L$ rungs, consisting of the spin singlets and the lowest spin triplets, i.e., 
$\{|S \rangle, |T_{-} \rangle\}$ on the rungs, the spin-orbit coupled Fermi-Hubbard model reduces to an effective one-dimensional XXZ Heisenberg model in a longitudinal and a transverse effective  magnetic field (see~\ref{sec:hamiltonian}, \cite{Zhang2025UniversalControlST, farina2025site}). Denoting
$|\tilde{\uparrow} \rangle = |S \rangle$ and
$|\tilde{\downarrow} \rangle = |T_{-} \rangle$, the Hamiltonian can be written as

\begin{equation}
\label{eq:hamiltonian_main}
\begin{aligned}
H &= 
J_{\parallel} \sum_{i}^{L-1} \left( \tilde{S}_i^x \tilde{S}_{i+1}^x
+ \tilde{S}_i^y \tilde{S}_{i+1}^y \right)
+ \frac{J_{\parallel}}{2} \sum_{i}^{L-1} \tilde{S}_i^z \tilde{S}_{i+1}^z
\\
&\quad+
\sum_i^L \left( h_{z, i} - J_{\perp} - \frac{J_{\parallel}}{2} \right)\tilde{S}_i^z
+\frac{D}{2\sqrt{2}}\sum_{i}^L \tilde{S}_i^y .
\end{aligned}
\end{equation}

Here, $J_{\perp}$ and $J_{\parallel}$ are the exchange couplings along the rungs and legs, respectively; $h_{z,i} = \mu_B \overline{g_i} B^z$ is the average Zeeman energy, $\mu_B$  is Bohr magneton, and $\overline{g_i}$ the average $g$-factor of the two dots on rung $i$. 
$\tilde{S}_i^\alpha=\frac{1}{2}\tilde{\sigma}_i^\alpha$, where $\alpha \in \{x,y,z\}$ and $\tilde{\sigma}_i^\alpha$ are the Pauli matrices. 
The Dzyaloshinskii-Moriya term $D$ captures the  effect of SOI. We note here that effects of anisotropic exchange interactions are not taken into account~\cite{saez2025exchange}, and the current Hamiltonian description is sufficient to interpret the experiment.

We first consider the limit of negligible leg coupling, $J_{\parallel}\approx 0$, assume equal average Zeeman energies for each rung, and temporarily neglect the SOI term. In this regime, the ladder reduces to a collection of independent rung dimers, i.e. isolated double-dot pairs along the rungs of the ladder. Each rung has a singlet ground state for $J_{\perp}>h_z$ and a polarized triplet ground state for $J_{\perp}<h_z$, with a singlet-triplet level crossing at $J_{\perp}=h_z$. In the thermodynamic limit, the ground state switches abruptly from a product of rung singlets, with $M=0$, to a product of polarized triplets, with $M=1$. Here $M=\frac{1}{L}\sum_{i=1}^{L}m_i$, with $m_i\in\{0,1\}$ being the local triplet occupation of rung $i$, refers to the triplet density. This level crossing thus produces a first-order quantum phase transition in this system, at $J_\parallel = 0$ and as as $J_\perp$ is swept across the critical value $h_z$.

For finite leg coupling, $J_{\parallel}\neq 0$, the local triplet excitations become mobile quasiparticles, which are called triplons~\cite{sachdev2000quantum, farina2025site}. Unlike the total magnetization, the rung-triplet number per site is then not an exact conserved quantity and may fluctuate. In the thermodynamic limit, a new phase emerges, along with two critical points, $J_{\perp, C_1}$ and $J_{\perp, C_2}$, as illustrated in Fig.~\ref{fig:fig1}(a,b). Below $J_{\perp, C_1}$ the system is in a rung-singlet phase (RS), while above $J_{\perp, C_2}$ it is in a fully polarized phase (FP). Both phases are gapped and lack long-range correlations or long-range order. Between the two critical points, the excitation gap closes and the system enters a gapless Tomonaga--Luttinger liquid (TLL) regime. The width of the gapless phase scales with the triplon bandwidth and is proportional to $J_{\parallel}$, i.e.
($J_{\perp, C_2}-J_{\perp, C_1})\propto J_{\parallel}$~\cite{Giamarchi2008BECMagneticInsulators}.
The triplet-number variance, $\mathrm{Var}(M)=\langle M^2\rangle-\langle M\rangle^2$,
is expected to peak near the critical points, where fluctuations in the triplon density are enhanced.

The critical boundary lines are determined by the energy costs of creating the lowest-energy excitations above the two limiting ground states: RS and FP. In the RS phase, this corresponds to injecting a single delocalized triplon, whereas in the FP phase, it corresponds to injecting a single delocalized singlon (Sec.~\ref{sec:slopestheory}, Ref.~\cite{CoupledLaddersMagneticField}). Near the lower boundary ($C_1$),
injecting a triplon into the singlet background lowers the system's energy
with increasing magnetic field. Conversely, near the upper boundary ($C_2$),
removing a triplon (or equivalently adding a singlon to) from the fully polarized state lowers the energy with
decreasing field. Because these processes carry opposite Zeeman shifts, the
critical lines exhibit opposite slopes in the phase diagram. 
Within the mean-field approximation, this yields the following phase boundaries:
\begin{equation}
(h_z-J_{\perp})_{C_1}=-J_{\parallel},
\qquad
(h_z-J_{\perp})_{C_2}=2J_{\parallel}.
\end{equation}

For a finite ladder with $L$ rungs, the magnetization evolves between these critical points through a discrete sequence of finite-size steps rather than continuously. In the limit of vanishing $U(1)-$
symmetry-breaking terms, these steps correspond approximately to successive changes in the triplon number by one. As $L$ increases, the steps become progressively denser, recovering the continuous magnetization curve of the thermodynamic limit. The corresponding finite-size level crossings converge toward the critical fields delimiting the intermediate phase.

We refer to this regime as a canted antiferromagnet (CAFM)~\cite{Giamarchi2008BECMagneticInsulators}, as it is defined by a finite magnetization along the applied field coexisting with strong quasi-long-range transverse antiferromagnetic correlations. In an idealized higher-dimensional system with true long-range order, this regime would correspond to spontaneous breaking of the continuous $O(2)$ symmetry in the transverse plane, giving rise to a gapless Goldstone mode. In contrast, SOI pins the ordering direction, resulting in a weakly gapped pseudo-Goldstone mode (see Supplementary Section~\ref{sec:BEC}).

\section{Experimental realization of quantum phase crossover in the small leg coupling regime}

In this work we perform experiments on a finite-size system in the presence of SOI, with $L = 4$  for the weak coupling regime and with $L = 3$ when reconstructing the full phase diagram. The device used in this experiment is a $2\times4$ quantum-dot ladder, fabricated in a Ge/SiGe heterostructure~\cite{Lodari2021LowPercolationDensity} and previously used in~\cite{Zhang2025UniversalControlST}. Gate voltages applied to the plunger gates electrostatically define the quantum dots, and the interdot coupling is controlled via the barrier gates (Fig.~1 (c)). Each rung of the ladder is treated as a double quantum dot. 

To measure the triplet number on each rung, we implement the pulse sequence in Fig.~\ref{fig:fig2}(a), which defines the general operational protocol of the experiment (see Supplementary Section~\ref{sec:pulse}). This protocol combines digital quantum gates during readout with analog quantum simulation techniques to prepare the ground state of Eq.~\ref{eq:hamiltonian_main}~\cite{farina2025site}. All operations are carried out pairwise along the rungs of the ladder. Initially, each rung dimer is initialized into a singlet state deep in the (0,2)/(2,0) charge configuration. To probe the ground state of the system, all rung dimers are simultaneously ramped to the operation point in the $(1,1)$ charge configuration. The ramp speed is chosen to yield adiabatic evolution through the $S-T_{-}$ anticrossing~\cite{Shevchenko2010,Kim2024AdiabaticStatePreparationIsing, farina2025site} (Supplementary Section~\ref{sec:ramp}) so the system reaches its ground state. Readout is performed via Pauli spin blockade and simultaneous charge detection~\cite{jirovec2021singletqubit}. This procedure effectively projects the system onto the singlet-triplet basis and freezes the state evolution. The spin state of pairs 1, 2 and 4 is read out through the respective sensors. For readout of pair 3, we obtain the highest fidelity by swapping its spin state with that of pair 2, followed by a second readout of pair 2~\cite{Zhang2025UniversalControlST}. This scheme allows us to extract in every experimental shot four bits of information, one for each rung.

We first study the regime of small but finite $J_{||} \ll h_z$, holding the Zeeman energy $h_z$ fixed and controlling the phase through $J_{\perp}/h_z$. Throughout the experiment we apply an external in-plane magnetic field of $5~\mathrm{mT}$. With rung $\overline g$-factors $0.33$ - $0.37$~\cite{Zhang2025UniversalControlST}, this corresponds to $h_z \sim 25~\mathrm{MHz}$. With careful characterization of the leg exchange couplings, we set $J_{\parallel} \approx 1~\mathrm{MHz}$ (Supplementary Section~\ref{sec:Jpar}). 

Fig.~\ref{fig:fig2}(b) shows the measured triplet number averaged over all rungs, along with its variance, when performing the pulse protocol of Fig.~\ref{fig:fig2}(a). The plotted values are obtained after processing the experimental data by accounting for readout errors (See Supplementary Section~\ref{sec:SPAM}) and applying a maximum-likelihood estimation (MLE) procedure to extract the triplet occupations per rung, $m_i$ (Supplementary Section~\ref{sec:MLE})~\cite{geller2020rigorous}. The horizontal axis uses the average quantity $\overline{(h_z - J_{\perp})}$ across all four rungs (Supplementary Section~\ref{sec:Jperp}). The point $\overline{(h_z - J_{\perp})} = 0$ corresponds to the critical point for decoupled rungs. For negative values of $\overline{(h_z - J_{\perp})}$, where exchange dominates, we record a low triplet density, consistent with a RS phase. For positive values, the Zeeman energy dominates, and we detect a large number of triplets, consistent with a FP phase. Finally, the measured variance of the triplet density exhibits a maximum near $\overline{h_z - J_{\perp}} = 0$, consistent with a cross-over between two phases. All these observed features are also seen in numerical simulations of the Fermi-Hubbard Hamiltonian in Eq.~\ref{Eq:mainH1}, which underlies the effective spin Hamiltonian of Eq.~\ref{eq:hamiltonian_main} (Supplementary Sections~\ref{sec:hamiltonian}). The microscopic Hamiltonian parameters, in particular the on-site interaction energy, the hopping terms, the local $g$-factors and the SOI terms, are extracted from independent spectroscopic experiments by Hamiltonian learning methods (Supplementary Section~\ref{sec:theoreticalfit}). The simulation results are overlaid on top of the data and show good overall agreement with the data (except for a larger than expected measured variance at negative $\overline{h_z - J_{\perp}}$). 

We further examine the structure of the crossover by using the same data set to plot the probability for each triplet number~\cite{TripletProfileQDs}. As discussed in the previous section, finite-size plateaus in the triplet density are smoothed by SOI. However, the probabilities for each triplet number display individual resolved peaks, as seen in Fig.~\ref{fig:fig2}(c). 

Leveraging individual readout of each rung in every shot of the experiment, we also map the probability distribution over all 16 possible singlet and triplet outcomes on the rungs. Figure~\ref{fig:fig2}(d) shows the evolution of this distribution as $J_\perp$ is swept through the phase crossover. In the RS phase, the state $|SSSS\rangle$ dominates, while in the FP phase the system is dominated by $|TTTT\rangle$ (here $T \equiv T_{-}$). In the intermediate regime, we observe a gradual increase in the number of triplets. Furthermore, we observe that all possible one-, two- and three-triplet probabilities exhibit a peak in the intermediate (CAFM) phase. This shows that the triplets present in each case do not localize onto a single rung but are distributed over all sites, as expected theoretically (see the dotted lines which are numerical results). The triplet number at each point is therefore a collective property of the system. Furthermore, we observe in the experiment a tendency for triplets to repel each other, in fact more strongly than is expected theoretically. The quantitative discrepancies may arise from several sources, including slight variations in the leg and rung couplings due to experimental drift and limitations of the calibration procedures.

\begin{figure*}[t]
    \centering
    \includegraphics[width=1.0\textwidth]{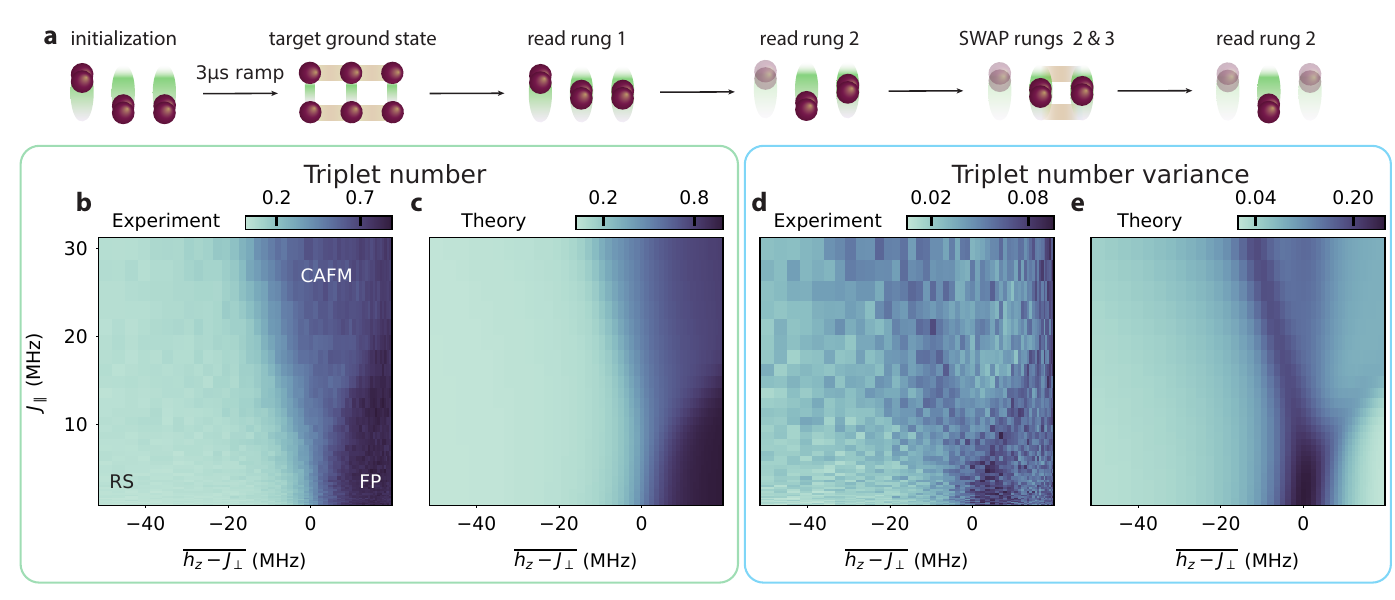}
    \caption{%
    \textbf{Phase diagram.}
    (a) Pulse sequence used to acquire the experimental data: initialization deep in the (0,2)/(2,0) charge configurations of the rungs $\rightarrow$ adiabatic evolution following the ground state $\rightarrow$ readout of dimer~1 $\rightarrow$ readout of dimer~2 $\rightarrow$ SWAP operation between dimers~2 and~3 $\rightarrow$ readout of dimer~2, carrying the state of dimer~3.
    (b) Phase diagram showing the measured average triplet number per rung as $J_{\perp}$ and $J_{\parallel}$ are swept. 
    (c) Corresponding theoretical phase diagram obtained by solving the Hamiltonian in Eq.~\ref{Eq:mainH1} using realistic experimental parameters.
    (d) Measured averaged variance of the triplet number per rung as $J_{\perp}$ and $J_{\parallel}$ are swept. 
    (e) Theoretical triplet number variance, similar to (c).
    }
    \label{fig:fig3}
\end{figure*}

Throughout this discussion, we have assumed that the ground state of the system is reached for all values of $\overline{(h_z - J_{\perp})}$. We here examine this assumption more quantitatively. Every pair of the system is initialized deep in the $(0,2)/(2,0)$ configuration, where the \(S\)--\(T_-\) energy separation is set by the double dot detuning and is on the order of \(100~\mathrm{GHz}\) \cite{Hendrickx2020FastTwoQubitLogic}, much larger than the lattice temperature, which is expected to be of the order of 10 mK base temperature of the dilution refrigerator. From there, the detuning and interdot tunnel coupling are varied adiabatically, taking the system to a condition where the gap is orders of magnitude smaller. In the FP (RS) phase, the Zeeman (exchange) energy of roughly \(25~\mathrm{MHz}\) (\(55~\mathrm{MHz}\)) sets the gap energy scale. In the CAFM phase, the gap is expected to be even smaller, see Fig.~\ref{fig:fig1}(a). For a perfectly adiabatic transition from $(0,2)/(2,0)$, the effective temperature reached is smaller than the lattice temperature by the same ratio as the gap energy scales, which is more than three orders of magnitude. The system could thus potentially reach an effective temperature below 10 $\mu$K after the adiabatic transition. We find empirically that the best agreement between the numerical results and the experimental data is reached using a slightly larger temperature of \(50~\mu\mathrm{K}\) in the simulations  (Supplementary Section~\ref{sec:ramp}), which is reasonable given imperfect adiabaticity.

\section{Probing Quantum Phase Crossover for a large leg coupling}

For the experimental exploration of the phase diagram whereby also $J_\parallel$ is varied, we use the three leftmost rungs out of the four rungs of the ladder. Furthermore, we focus on the case of strongly asymmetric leg couplings, where $J_\parallel$ is swept only for the top leg of the ladder. As shown in Fig.~3(a), the pulse scheme is nearly identical to the one used for four rungs, and also calibration of the $x$-axis and MLE data processing is performed in the same way as in the four-rung experiments. Calibration of the horizontal exchange couplings is carried out via a series of sequential measurements of the energy splitting between the singlet and triplet $T_0$ states as a function of the external magnetic field (Supplementary Section~\ref{sec:Jpar}). The magnetic field is still held at \(5\,\mathrm{mT}\).

Together, these calibration procedures allow us to construct the experimental phase diagram shown in Fig.~3(b). The $y$-axis represents the average $J_\parallel$ along the upper leg, and the color scale represents the triplet density (the triplet number per rung averaged per number of rungs). We observe three distinct regions, consistent with theoretical expectations: a low-triplet number region identified with the RS phase, an intermediate-triplet number CAFM phase, and a FP phase with maximal triplet number. Figure~3(d) shows the triplet number variance for the same data set as in Fig.~3(b). We observe lines of peaked variance, which are typically identified with phase crossovers. Indeed, the position of these lines coincides with the boundaries of the middle (CAFM) region.

We can also identify two important features that are in line with the theoretical prediction: (i) The CAFM phase expands with increasing $J_{\parallel}$ toward both positive and negative values along the $x$-axis. It is worth noting that the expansion of the CAFM region becomes visible only when $J_{\parallel}$ is sufficiently large, around 10 MHz, to overcome the visual smearing of the plateaus caused by SOI. (ii) The two slopes separating the CAFM phase from the neighboring phases are different. In particular, the absolute value of the experimentally extracted slope associated with the RS-to-CAFM crossover (Supplementary Section~\ref{sec:slopes}) is approximately a factor of 2 smaller than that associated with the CAFM-to-FP crossover~\cite{CoupledLaddersMagneticField}, consistent with the theoretical value of 2. 

\begin{figure*}[t]
    \centering
    \includegraphics[width=\textwidth]{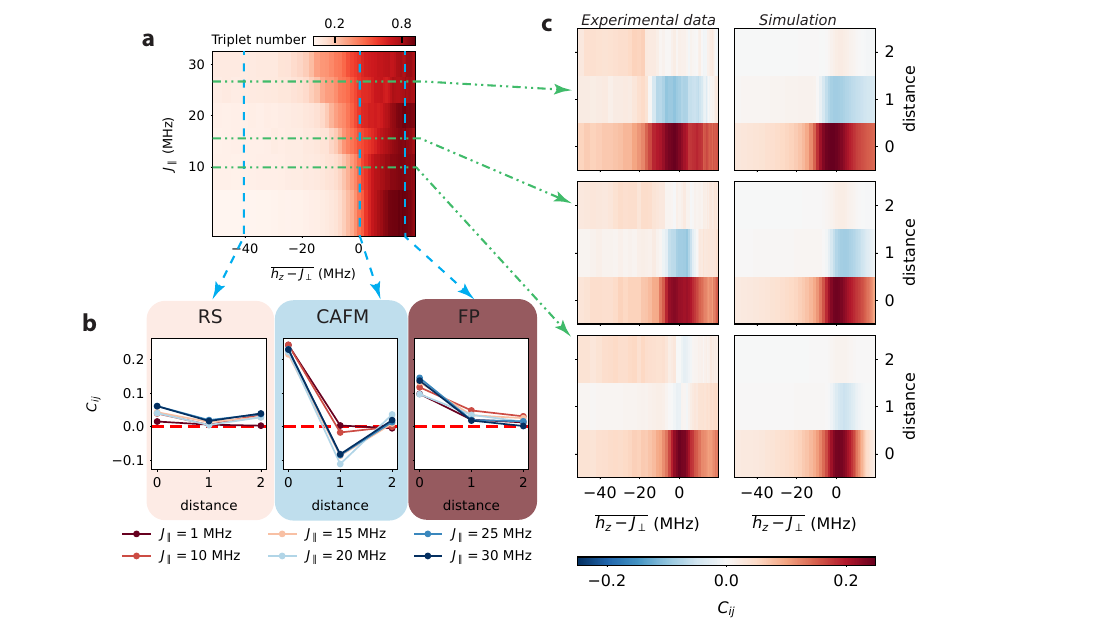}
    \caption{%
    \textbf{Connected correlators.}
    (a) Phase diagram reconstructed from one-dimensional scans for \(J_{\parallel} \in \{1,10,15,20,25,30\}\,\mathrm{MHz}\), using a per-scan SPAM matrix.
    (b) Line cuts for each sampled value of \(J_{\parallel}\), showing the connected-correlator amplitude deep within each phase as a function of distance $j-i$. Distance $0$ denotes the autocorrelator, distance $1$ the nearest neighbour, and distance $2$ the next-nearest neighbor correlator. The RS and FP phases show minimal residual correlations for all values of \(J_{\parallel}\), whereas the CAFM phase develops anti-correlations as \(J_{\parallel}\) increases, reflecting repulsive interactions between neighboring triplons.
    (c) Horizontal line cuts of the connected correlator extracted from the phase diagram at \(J_{\parallel} \in \{10,15,25\}\,\mathrm{MHz}\), shown for each distance and each value along the \(x\)-axis, and compared with numerical simulations. We observe an expansion of the region exhibiting anti-correlations at distance $1$ with increasing \(J_{\parallel}\), in good visual agreement with theory.
    }
    \label{fig:fig4}
\end{figure*}

Lastly, we highlight an additional feature at the right edge of the phase diagram, where $J_{\perp}$ tends to zero ($J_{\perp}$ cannot go negative). As $J_{\parallel}$ is increased, the energy of the fully polarized state in the FP phase eventually becomes equal to that of the state with a single flipped spin, inside the CAFM phase. This crossing defines a critical leg coupling, $J_{\parallel,C} = \tfrac{2}{3} h_z$. For $J_{\parallel} > J_{\parallel,C}$, the total triplet number therefore cannot reach its maximal value of $3$, a behavior observed experimentally at the rightmost edge of the phase diagram. 

We directly compare the experimental results with the theoretically predicted phase diagram in terms of triplet number (Fig.~\ref{fig:fig3}(c)) and the corresponding triplet number variance (Fig.~\ref{fig:fig3}(e)). These theoretical plots are obtained by solving the Hamiltonian in Eq.~\ref{Eq:mainH1} of Supplementary Section~\ref{sec:theoreticalfit} using realistic experimental parameters. Qualitatively, we find strong agreement between theory and experiment for both the triplet number and the triplet number variance. The theoretical plots, however, do not include readout errors and leg SOI and therefore allow for a better resolution of the plateaus within the CAFM region, with lighter shading corresponding to an averaged triplet number of $1/3$ and darker shading corresponding to $2/3$. The transition between these two regions within the CAFM phase is less sharp than the transitions separating the three distinct phases, and is therefore harder to resolve experimentally, although it is visible in the data shown in Supplementary Section~\ref{sec:PD_from_corr}. Moreover, the CAFM phase is particularly sensitive to the presence of SOI, and including a small inhomogeneity in the leg SOI is found theoretically to quickly smear out the transitions inside the CAFM phase.

\section{Correlations}

In order to further characterize the three phases, we access the connected longitudinal correlator,
\begin{equation}
C_{ij} = \langle m_i m_j \rangle - \langle m_i \rangle \langle m_j \rangle \,.
\label{eq:correlator}
\end{equation}

In the absence of SOI, the CAFM phase exhibits commensurate antiferromagnetic correlations with wave vector $Q=\pi$. In this case, the correlator of Eq.~\ref{eq:correlator} is expected to decay algebraically with distance and to exhibit an alternating sign, corresponding to quasi-long-range order~\cite{Mermin, Brown2017SpinImbalance,salomon2019direct}. In the presence of SOI, assuming $J_{\parallel} \gg D$, the wave number is shifted to $Q=\pi+q_0$, with $q_0 \propto -D/(\sqrt{2}J_{\parallel})$. Increasing $J_{\parallel}$ thus suppresses SOI effects and enhances antiferromagnetic order (See~\ref{sec:bethe}). In contrast, both the RS and FP phases lack correlations and therefore $C_{ij}$ is expected to vanish.

We now record and process data using an individual State Preparation and Measurement (SPAM) matrix for each value of $J_{\parallel}$. This provides additional precision, reduces calibration-induced errors, and allows us to extract the observables at each point more accurately. Though we perform this procedure for only six values of $J_\parallel$ given the additional overhead, we can reconstruct a coarse phase diagram from these data as well, shown for illustration purposes in Fig.~\ref{fig:fig4}(a). Fig.~\ref{fig:fig4} (b) shows the correlation versus distance for three values of $\overline{h_z - J_{\perp}}$, corresponding to each of the three phases, indicated by the blue dashed lines in Fig.~\ref{fig:fig4}(a). Fig.~\ref{fig:fig4}(c) shows the correlation versus distance at three different values of $J_{\parallel}$, indicated by the green dashed-dotted lines in Fig.~\ref{fig:fig4}(a). 

We observe no appreciable inter-site correlations in neither the RS or the FP phase, in agreement with theoretical expectations.  Autocorrelations ($j=i$) are still finite since they reflect the variance. In the CAFM regime, anti-correlations ($C_{ij} <0$) between neighboring rungs are present in experiment for $J_{\parallel} \geq 20$ MHz. Larger system sizes would be needed to probe the predicted power-law decay in the correlations. For smaller $J_\parallel$ and considering a crude estimate of $D\approx 5$ MHz, we no longer expect a negative correlation at  distance 1, i.e. $j=i+1$. 
Moreover, in Fig.~\ref{fig:fig4}(c) we can see a widening of the region with negative correlation at distance 1 with increasing $J_{\parallel}$, in agreement with the structure of the phase diagram and the interpretation of the middle phase as a CAFM phase.

We stress that quantum dot systems provide access beyond local observables. Beyond the triplet probabilities on individual rungs, we can probe the joint probability distribution of the underlying spin configurations and provide direct access to triplet–triplet correlators, i.e. four-point correlation functions, which are exceptionally difficult to measure in conventional solid-state platforms. In principle, using digital quantum operations acting on the $\{|S\rangle,|T_-\rangle\}$ subspace would allow us to probe also the transverse correlators, which would provide conclusive evidence of the CAFM phase (Supplementary Section~\ref{sec:BEC}).

\section{Conclusion}
To summarize, we have explored quantum magnetism in a quantum dot ladder by studying ground state properties at various Hamiltonian settings. Calibration methods allowed us to controllably vary the spin exchange couplings along the legs and the rungs of the ladder, and we extracted the on-site interaction energy and spin-orbit interaction terms through Hamiltonian learning techniques. Despite the small system size, our results -- namely magnetization measurements and variances thereof, and correlation functions capturing long-range order -- show good agreement with theoretical predictions for the ground state of the Heisenberg spin ladder as a function of rung and leg couplings, transitioning from a rung singlet phase to a canted antiferromagnetic phase and to a fully polarized phase. The finite system size furthermore results in a finite gap in the CAFM regime. Due to the presence of spin-orbit interaction, sharp quantum phase transitions are broadened into phase crossovers.

Placing this work in a broader context, the phase cross-overs can be described within the framework of Bose--Einstein condensation of hard-core bosons~\cite{Giamarchi2008BECMagneticInsulators,Zapf2014BECQuantumMagnets}. In this mapping, a triplon represents a hard-core boson and a singlet corresponds to the absence of a boson. The hard-core constraint arises from the fact that each rung can host at most one triplonic excitation. Within this picture, the rung-singlet phase corresponds to the vacuum of triplons, the CAFM phase corresponds to a Bose-Einstein condensate of triplons, and the fully polarized phase corresponds to a Mott-insulating state with one triplon per rung. With more ladders weakly coupled together, this platform can provide a direct experimental realization of the theoretical proposal of Ref.~\cite{Giamarchi2008BECMagneticInsulators} to investigate Bose-Einstein condensation in solids.

Although this work is restricted to the half-filled Fermi–Hubbard model, doped Fermi–Hubbard ladders can also be prepared, enabling studies of magnetic polarons formed by doped charges moving in an antiferromagnetic background~\cite{Koepsell2019MagneticPolarons}. Their motion and correlations are closely tied to pairing tendencies in doped systems~\cite{arovas_hubbard_2022}, while higher-order spin–charge correlations can reveal their internal structure. The ability to measure higher-order spin-correlation functions may also enable studies of distribution functions and hyperscaling near continuous phase transitions~\cite{AjiGoldenfeld2001Fluctuations}. Dynamical experiments could further be used to investigate magnetic excitations~\cite{Lorenzana1995PhononAssisted, farina2025site}, while correlation-function measurements in driven systems could provide access to the nature of excited states. 
More broadly, the Hamiltonian learning approach demonstrated here, which enables the accurate determination of Hamiltonian parameters, including SOI, is crucial for developing algorithms that leverage quantum-dot-based platforms to solve practical problems~\cite{SpinQubits,dijkema2026simultaneous}, such as simulating nuclear spin dynamics for NMR inference, where precise knowledge of the underlying spin Hamiltonian is required to faithfully reproduce the target dynamics~\cite{Sels2020QuantumApproxBayesianNMR,Seetharam2023DigitalNMR,Fratus2025QuantumClassicalNMR,OBrien2022NMRMolecularStructure}. Future studies of the rich phases that emerge in the presence of strong SOC could similarly benefit from these techniques~\cite{Chou2025SpinLadderQuantumSimulators}.

Altogether, our results provide a compelling motivation for the exploration of larger semiconductor quantum-dot arrays with various lattice geometries~\cite{dijkema2026simultaneous,HRLQuantumTeam2026Digital,ChargeFrus2013,Kiczynski2022EngineeringTopologicalStates} as a platform for quantum simulations of many-body physics, frustrated magnetism, and exotic phases of matter.

\section*{Acknowledgments}
We thank Stefano Reale, Maximilian Rimbach-Russ, Stefano Bosco, and other members of the Vandersypen, Veldhorst, Scappucci, Rimbach-Russ and Bosco groups for nice discussions and kind help. We acknowledge
S. L. de Snoo’s and D. Bijl's help in software development and technical support by
O. Benningshof, J. D. Mensingh, T. Orton, R. Schouten,
R. Vermeulen, R. Birnholtz, E. Van der Wiel, B. Otto,
D. Brinkmans. This work was funded by
an Advanced Grant from the European Research Council (ERC) under
the European Union’s Horizon 2020 research (882848) and NWO Spinoza prize awarded to
L.M.K.V. by the Netherlands Organisation for Scientific
Research (NWO/OCW). U.B. and E.D. acknowledges financial support from the Swiss National Science Foundation
(SNSF) through project No.~200021\_212899, the Swiss State Secretariat for
Education, Research and Innovation (SERI) under contract No.~UeM019-1, and
NCCR SPIN, a National Centre of Competence in Research funded by the Swiss
National Science Foundation (grant No.~225153). U.B. is also grateful for the financial support of the IBM Quantum Researcher Program.

\section*{Author contributions}
The experiments and data analysis were performed by E.M. and X.Z. Theoretical calculations were performed by U.B. E.M., X.Z., U.B., P.C.F., D.J., E.D., and L.M.K.V. contributed to interpretation of the results. A. N.-K. and S.B. contributed to preliminary theoretical discussions. S.O. fabricated the device under the supervision of M.V. G.S. developed and provided the heterostructure. E.D., L.M.K.V. and X.Z. conceived the experiment. The manuscript was written by E.M., X.Z., U.B. and L.M.K.V., with input from all co-authors. L.M.K.V. and E.D. supervised the project.

\section*{Data availability}
The data and analysis supporting this work are available on Zenodo via  https://doi.org/10.5281/zenodo.21983863.

The code used to generate the numerical results presented in this study is available from ED upon reasonable request.

\section*{Competing interests}
The authors declare no competing interests.

\bibliographystyle{naturemag}
\bibliography{main}
\clearpage
\onecolumngrid
\appendix
\section*{Supplemental Material}
\setcounter{subsection}{0}
\renewcommand{\thesubsection}{S\arabic{subsection}}
\setcounter{subsubsection}{0}
\renewcommand{\thesubsubsection}{S\arabic{subsection}.\arabic{subsubsection}}
\makeatletter
\renewcommand{\p@subsubsection}{}
\@addtoreset{figure}{subsection}
\@addtoreset{figure}{subsubsection}
\makeatother
\newif\ifsubsubsecmulti
\subsubsecmultifalse
\setcounter{figure}{0}
\renewcommand{\thefigure}{%
  \thesubsection%
  \ifnum\value{subsubsection}>0%
    .\arabic{subsubsection}%
  \fi%
  \ifsubsubsecmulti
    .\arabic{figure}%
  \fi%
}
\pretocmd{\subsection}{\setcounter{figure}{0}}{}{}
\pretocmd{\subsubsection}{\setcounter{figure}{0}\subsubsecmultifalse}{}{}
\raggedbottom

Supplementary is organized as follows:
\begin{itemize}
  \item \textbf{Sec.~\ref{sec:protocol}} --- General Operational Protocol
  \item \textbf{Sec.~\ref{sec:virtual}} --- Virtual gate matrix
  \item \textbf{Sec.~\ref{sec:errors}} --- Readout Error Mitigation
  \item \textbf{Sec.~\ref{sec:Gaussian}} --- Comparison to Gaussian MLE
  \item \textbf{Sec.~\ref{sec:Jperp}} --- Calibration of the rung exchange
  \item \textbf{Sec.~\ref{sec:Jpar}} --- $J_{\parallel}$ calibration
  \item \textbf{Sec.~\ref{sec:theory}} --- Theoretical model
  \item \textbf{Sec.~\ref{sec:theoreticalfit}} --- Theoretical fit and comparison with the experimental data
  \item \textbf{Sec.~\ref{sec:BEC}} --- Bose--Hubbard mapping, transverse correlator, and the canted antiferromagnet
  \item \textbf{Sec.~\ref{sec:slopes}} --- Slope extraction from phase diagram
  \item \textbf{Sec.~\ref{sec:PD_from_corr}} --- Additional phase diagram data
\end{itemize}

\clearpage

\subsection{General Operational Protocol}
\label{sec:protocol}

\subsubsection{Pulse sequence}
\label{sec:pulse}
To experimentally probe the observables discussed in this manuscript, we implemented the pulse sequence illustrated in Fig.~\ref{fig:fig2}(a), which defines the general operational protocol of the experiment. All operations are carried out pairwise along the rungs of the ladder, and we use both the bottom-left ($S_{\mathrm{BL}}$) and the bottom-right ($S_{\mathrm{BR}}$) sensor to measure the charge occupation of all the quantum dots. Each dot pair is initialized into its ground state, a singlet state, in the (0,2)/(2,0) double-dot charge configuration after a 50~$\mu$s waiting period. After that, all rung dimers are simultaneously ramped to the operation point in the (1,1) charge configuration with target $J_\perp, J_\parallel$ values. The ramp time is 3 $\mu s$, which is chosen to be adiabatic with respect to the $S-T_{-}$ anticrossing (of approx. $10~\mathrm{MHz}$), such that the ground state of the system at half filling (one hole per site) is reached~\cite{Kim2024AdiabaticStatePreparationIsing}. The 2--20~ns duration at the operation point is finite simply to ensure complete implementation of the sequence and to temporally fix the final ground state. Evolution of the prepared state due to coupling to the environment is expected to be negligible during this time ~\cite{Zhang2025UniversalControlST}. Immediately afterwards, readout of the outermost rung dimers (1 and 4) is performed via direct PSB, while dimers 2 and 3 are diabatically ramped to positions where the intra-dimer exchange interaction values are much larger than $\Delta ST_{-}$ and $h_z$. This procedure effectively projects the system onto the singlet-triplet basis and freezes the state evolution. After the readout of the outermost pairs, the voltages of rung dimer~1 are adjusted to enable direct PSB readout of dimer~2. Finally, an effective SWAP gate is applied to dimers~2 and~3; as a result, dimer~2 carries the state of dimer~3, and readout is performed in the same manner as in the previous step. Here, the choice of the SWAP gate for readout of rung 3 is motivated by its higher fidelity compared with the indirect PSB-based readout demonstrated in~\cite{Zhang2025UniversalControlST}. All readout integration times are set to 30~$\mu$s. This scheme allows us to extract information about the spin state on all rungs of the system in a single experimental shot.

For measurements of the phase diagram and correlations shown in Fig.~\ref{fig:fig3}, the pulsing scheme is nearly identical to that used in the four-rung case. The only difference is that, in the first readout step, we focus solely on dimer~1 rather than performing simultaneous readout of dimers~1 and~4. The remaining steps of the pulse sequence remain unchanged.

\subsubsection{Ramp time and effective temperature}
\label{sec:ramp}

\begin{figure}[h]
\centering
\includegraphics[width=1.0\textwidth]{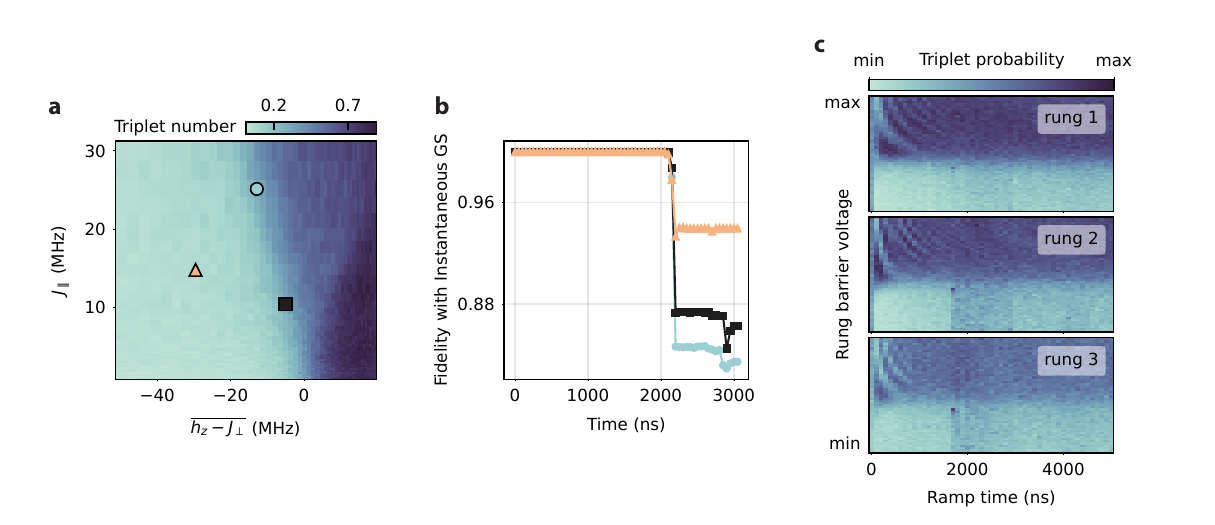}
\caption{%
\textbf{Instantaneous ground state and ramp time calibration.}
(a) Points indicate the positions where the instantaneous ground state was checked in (b).
(b) Overlap of the state at each point of the ramp with the instantaneous ground state, each line corresponding to a different final state as indicated by the individual markers in (a).
(c) Experimentally measured spin triplet probability as a function of ramp time and rung barrier voltage. We observe that a 3 $\mu$s ramp time is sufficient to avoid inducing oscillations involving excited states of the system.
}
\label{fig:ramp}
\end{figure}

We perform numerical simulations of the quantum dynamics of a six-site spinful fermionic ladder using exact diagonalization and Krylov-subspace time evolution within the QuSpin framework~\cite{weinberg2017quspin,weinberg2019quspin2}. The system is initialized in the many-body ground state obtained via Lanczos diagonalization of the Hamiltonian at the initial parameter point. The rung detuned onsite potential is chosen to be strongly negative such that it overcomes the interdot Hubbard repulsion $U$, thereby stabilizing an initial $S(2,0)$-type singlet configuration with two opposite-spin fermions localized in the minimum of the onsite potential. This state serves as the starting point for the subsequent ramp dynamics. The time evolution is then computed using a Krylov--Lanczos propagation scheme for a Hamiltonian with time-dependent tunnelling and detuning parameters, ramped linearly over a microsecond timescale to their final values, while interaction, Zeeman, and spin--orbit couplings remain fixed. To quantify the degree of adiabaticity, we compute the overlap between the time-evolved state and the instantaneous ground state (i.e., at each time step of the ramp) at discrete time intervals.

As shown in Fig.~\ref{fig:ramp} (a,b), the fidelity (defined as the overlap squared) remains typically close to $90\%$ throughout the ramp, indicating that the evolution closely follows the instantaneous ground-state manifold despite finite ramp speeds. The residual deviations from perfect adiabaticity manifest as weak excitations above the ground state. We captured these effects by modeling the final state as a thermal density matrix in the vicinity of the ground state, as this provides a convenient phenomenological framework for incorporating non-adiabatic excitations. We found that an effective temperature $T$ of 0.1 mK for 3 rungs and 0.05 mK for 4 rungs yields good agreement with experimentally measured observables.

Experimental probing of the ramp time that ensures adiabaticity is performed using a similar protocol to the measurement of the phase diagram, but with the ramp time varied. The measurements are shown in Fig.~\ref{fig:ramp}(c). One can notice that, for ramp times below 2 $\mu$s, some dynamics are still observed, but they fade as the ramp time approaches 3 $\mu$s, the value chosen for the experiment.

\clearpage

\subsection{Virtual gate matrix}
\label{sec:virtual}

Similar to Refs.~\cite{rao2025modular,Jirovec2025MitigationExchangeCrosstalk}, we used five layers of virtual gates to compensate for capacitive cross-talk. In the first layer, we compensate for the influence of the quantum dot ladder gates on the charge sensors. In the second layer, we orthogonalize the plunger gates, enabling independent control of the chemical potential of each quantum dot in the ladder. In the third layer, we normalize the charging energies of the quantum dots. In the fourth layer, we compensate for the crosstalk from the interdot barrier gates onto the quantum dot chemical potentials. Finally, in the fifth layer, we compensate for the crosstalk among the interdot barrier gates to achieve independent control of the exchange couplings, which is essential for simulating the Heisenberg model. The crosstalk is measured when the detunings of all double pairs are at zero.

\begin{figure}[b]
\centering
\includegraphics[width=\textwidth]{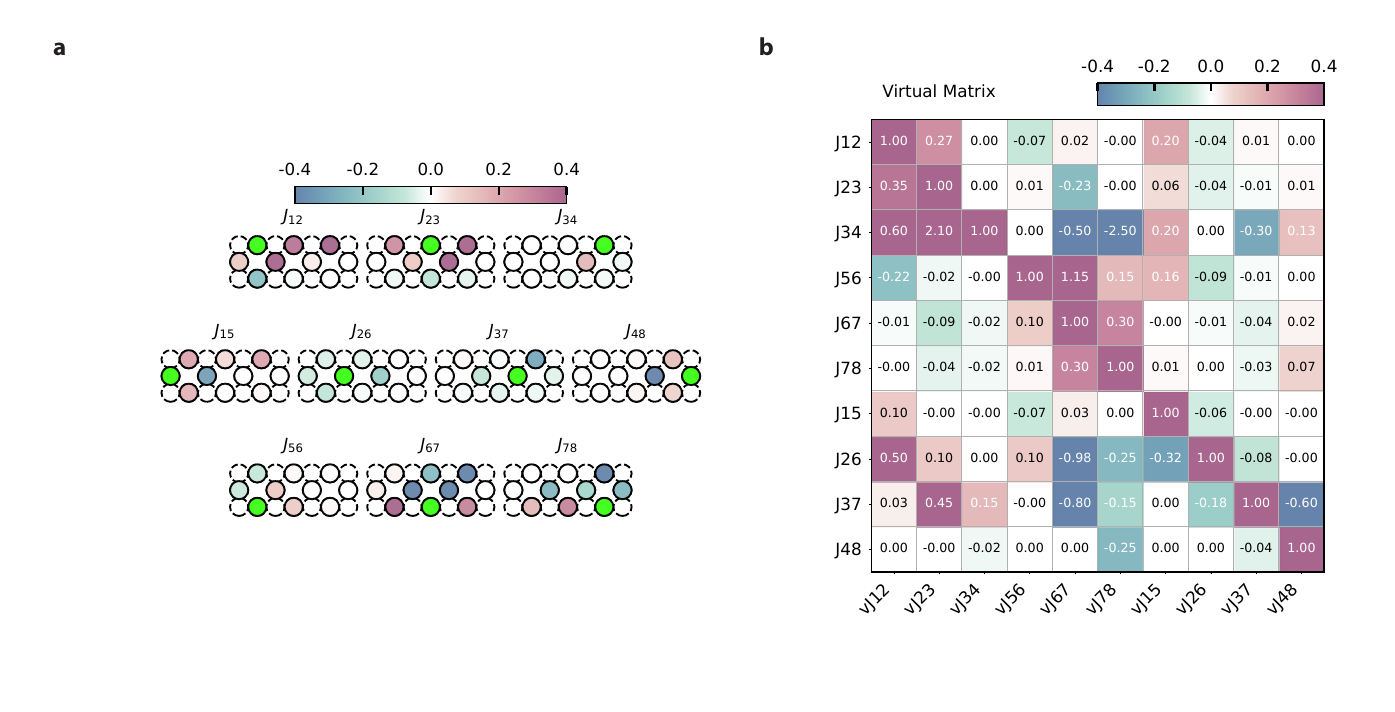}
\caption{%
\textbf{Virtual gate matrix.}
(a) Schematic illustration of the influence of the surrounding barrier gates on a selected exchange coupling. The color indicates by how much the voltage on the target barrier gate must be varied when the voltage applied to a neighbouring barrier gate (highlighted in bright green)  is changed by one unit. The cross-talk amplitudes are taken directly from the virtual gate matrix.
(b) Full virtual gate matrix capturing the cross-talk coefficients between barrier gates. We do not show the corresponding matrices for plunger–plunger or plunger–barrier cross-talk, as they are very similar to those reported in~\cite{Zhang2025UniversalControlST}. The gate voltage pulses $vJ_{ij}$ are constructed column-wise as linear combinations of the barrier voltages $J_{ij}$: for a given virtual gate (column), the barrier voltages (rows) are multiplied by the corresponding coefficients and summed. The exchange couplings $J_{ij}$ (i.e., the barrier voltages listed in the rows) are already virtualized with respect to the plunger gates.
}
\label{fig:virtual}
\end{figure}

Figure~\ref{fig:virtual}(a) shows a visual representation of the influence of the surrounding barrier gates on a given exchange interaction, which we aim to control using the barrier gate above it, highlighted in bright green. These plots are reconstructed from the virtual gate matrix shown in Fig.~\ref{fig:virtual}(b). The coefficients connecting the corresponding barriers are obtained under the assumption of linear dependencies between gate voltages: (a) by tracking the $ST_0$ oscillation frequency as a function of the surrounding barrier voltages for gates controlling $J_{\parallel}$, and (b) from the FFT of $ST_{-}$ Ramsey-style oscillations governed by gates controlling $J_{\perp}$ as a function of the surrounding barriers. One can notice a very strong cross-talk between $vJ_{23}$ and $J_{34}$ (here “$v$” denotes the composite virtualized gate voltage pulse that accounts for cross-talk via the coefficients of the virtual gate matrix). In particular, compensating for a pulse on $J_{34}$ requires applying a pulse 2.1 times larger on $vJ_{23}$, due to device-specific inhomogeneities that are not understood in detail. 
Considering the AWG output limits and attenuation in the setup, it was not possible to simultaneously reach large exchange pulses on both $vJ_{23}$ and $vJ_{34}$. As a result, the experiment was performed on four rungs in the case of small $J_{\parallel}$ and on three dimers when producing the phase diagram and correlation measurements, which require accessing large $J_{\parallel}$. We emphasize that this is a purely technical limitation and does not represent a fundamental constraint of the methods or concepts presented in this work.

\clearpage

\subsection{Readout Error Mitigation}
\label{sec:errors}

\subsubsection{Readout error mitigation}
\label{sec:SPAM}

\begin{figure}[b]
\centering
\includegraphics[width=\textwidth]{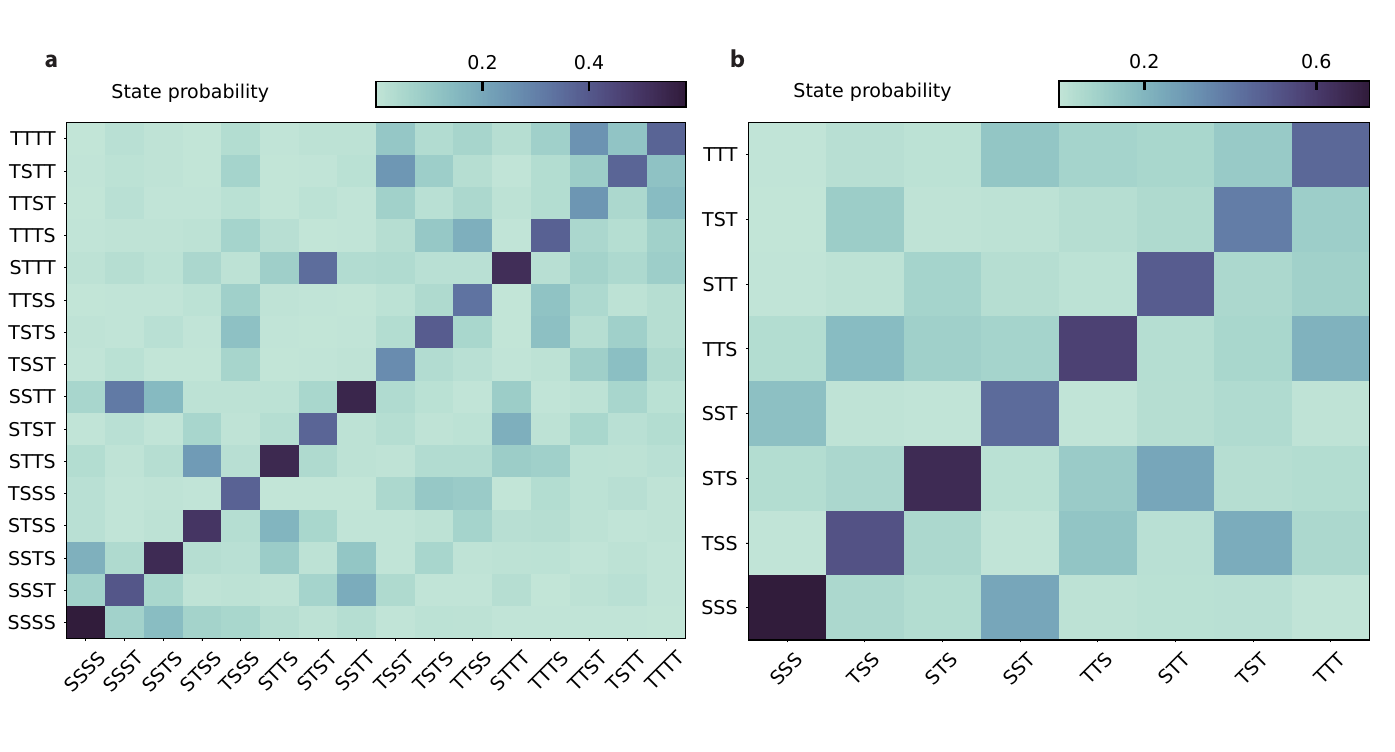}
\caption{%
\textbf{Readout error correction matrices.}
(a) SPAM matrix used for the four-rung measurements shown in Fig.~\ref{fig:fig2}.
(b) SPAM matrix used for the three-rung measurements shown in Fig.~\ref{fig:fig3} and Fig.~\ref{fig:fig4}.
Prepared states are shown along the $x$-axis, while columns correspond to the measured readout states.
}
\label{fig:SPAM}
\end{figure}

To mitigate readout errors, we apply a SPAM matrix to convert the measured probabilities to actual probabilities ( Fig.~\ref{fig:SPAM}), as discussed in~\cite{Zhang2025UniversalControlST}. The initialization error associated with the preparation of the local singlet ground state and the preparation of local triplet states via a $\pi$ pulse on the corresponding rung is relatively small. Consequently, the dominant contribution to SPAM errors in the experiment arises from the readout process. 

To quantify the readout errors, we perform calibration measurements in which all possible configurations of $S$ and $T_{-}$ (denoted simply as $T$ in Fig.~\ref{fig:SPAM}) are initialized and subsequently measured using the same simultaneous readout protocol as in the experiment. By comparing the prepared and measured states, we construct the SPAM matrix that to a good approximation characterizes the readout error probabilities. The matrix shows that the majority of the readout errors originate from rung~3. For example, the state $SSST$ is frequently misidentified as $SSTT$, and the state $STST$ is often read out as $STTT$. This is attributed to the finite fidelity of the SWAP gate between rungs~2 and~3.

These matrices are taken immediately before each experimental run. Here we show the exact matrices used for the subsequent maximum likelihood estimation (MLE) processing (Sec.~\ref{sec:MLE}) when producing Fig.~\ref{fig:fig2} and Fig.~\ref{fig:fig3}. For the correlation measurements shown in Fig.~\ref{fig:fig4}, individual SPAM matrices were taken for each value of $J_{\parallel}$, which are similar to the one shown in Fig.~\ref{fig:SPAM}(b).

\subsubsection{Maximum Likelihood Estimation}
\label{sec:MLE}

Direct inversion of the SPAM matrix does not capture possible drift between repetitions and may produce non-physical probabilities (negative or greater than one). Therefore, we use maximum likelihood estimation (MLE) to reconstruct the full probability distribution in the system given a measured SPAM matrix introduced in Section~\ref{sec:SPAM}. 

We discuss here the four-rung case, but the procedure is analogous for both the phase diagram and correlation functions. In Figure~\ref{fig:fig2}(b) and (c), each point on the $x$-axis (denoted by $x$) is measured with $N$ repetitions (in experiment $N = 300$). Measurements at different $x$ points are treated as statistically independent. The full experiment is repeated 60 times, producing 60 curves. For a given repetition $r$ and scan point $x$, the $N$ shots produce a count distribution over the 16 possible outcomes (all combinations of S and T for the four rungs). This defines a multinomial probability vector

\[
p_{r,x}
=
\big(
p_{r,x}^{(\mathrm{SSSS})},
p_{r,x}^{(\mathrm{SSST})},
\dots,
p_{r,x}^{(\mathrm{TTTT})}
\big),
\]
\[
\sum_{i=1}^{16} p_{r,x}^{(i)} = 1.
\]
We assume a fixed SPAM matrix $M$, calibrated independently prior to the experiment and treated as constant throughout data acquisition. The SPAM matrix acts on the true underlying distribution $p_{r,x}$ as

\[
q_{r,x} = M p_{r,x},
\]
where $q_{r,x}$ represents the expected measured distribution, considering measurement errors. Thus a state $\ell$ which occurs with probability $p_{r,x}^{(\ell)}$ is predicted to be observed as state $i$ with probability 

\[
q^{(i)}_{r,x}
=
\sum_{\ell=1}^{16}
M_{i\ell}\,
p^{(\ell)}_{r,x}.
\]

In the measured dataset, we denote $c^{(i)}_{r,x}$ the number of times outcome $i$ is observed in $N$ shots. Now the MLE works to find the $p_{r,x}^{(i)}$ which give the  $q_{r,x}^{(i)}$ that are most likely to generate the experimentally measured outcomes $c^{(i)}_{r,x}$, with the constraints $p_{r,x}^{(i)} \ge 0$ and $\sum_i p_{r,x}^{(i)} = 1$ for each scan point. The full multinomial likelihood for one repetition and scan point is

\[
\mathcal{L}(p_{r,x})
=
\frac{300!}{\prod_{i=1}^{16} c^{(i)}_{r,x}!}
\prod_{i=1}^{16}
\left(q^{(i)}_{r,x}\right)^{c^{(i)}_{r,x}}.
\]
This likelihood multiplies the predicted probability of each outcome, raised to the power corresponding to how many times that outcome occurs. The prefactor is the multinomial coefficient. Assuming independence across scan points, the total likelihood for repetition $r$ is

\[
\mathcal{L}(p_r)
=
\prod_{x}
\frac{300!}{\prod_{i=1}^{16} c^{(i)}_{r,x}!}
\prod_{i=1}^{16}
\left(q^{(i)}_{r,x}\right)^{c^{(i)}_{r,x}}.
\]
Because this likelihood is a product of many small probabilities, it becomes numerically unstable. We therefore minimize the negative log-likelihood, dropping constants independent of the parameters:

\[
-
\sum_{x,i}
c^{(i)}_{r,x}
\log\!\big((M p_{r,x})_i\big).
\]

The negative log-likelihood is minimized numerically under the constraints $p_{r,x}^{(i)} \ge 0$ and $\sum_i p_{r,x}^{(i)} = 1$ for each scan point. To ensure numerical stability, predicted probabilities are bounded away from zero during evaluation of the logarithm. This optimization finds the underlying probability curves that are most consistent with the experimental data per scan point and for each repetition separately.

\begin{figure}[t]
\centering
\includegraphics[width=\textwidth]{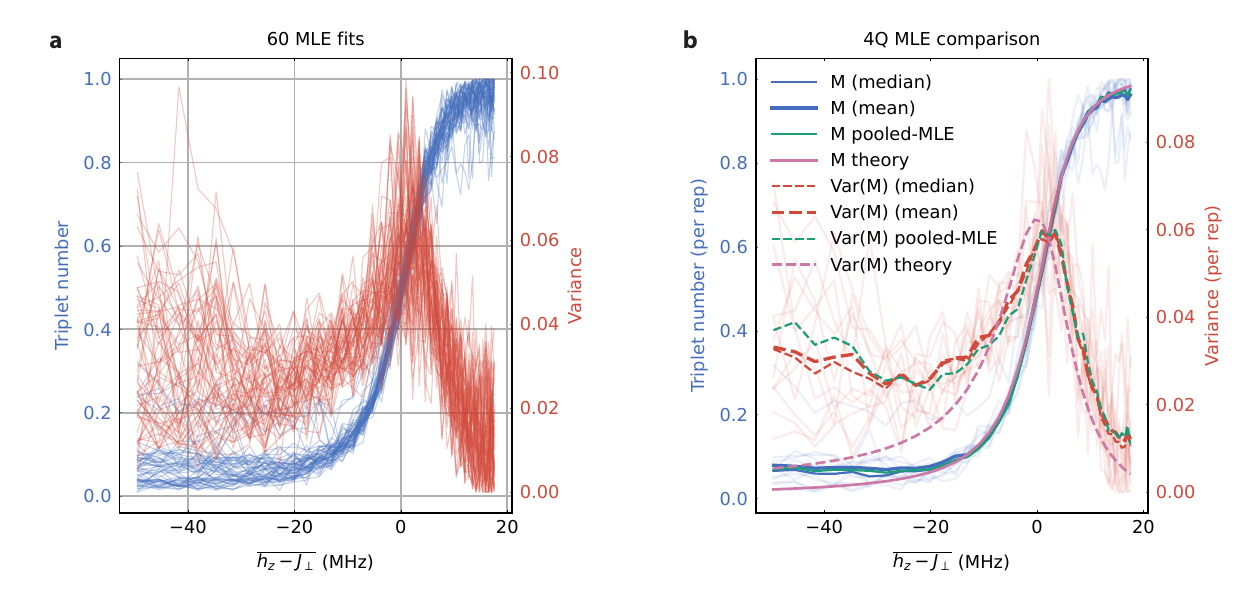}

\caption{%
\textbf{MLE comparison.}
(a) The blue lines show the triplet number resulting from performing 60 independent multinomial MLE fits. Each line corresponds to an independent repetition of the experiment. The red lines show the corresponding variance. The curves are constructed by fitting for each $x$ the distribution of counts over the 16 possible states (combinations of singlet and triplet configurations on four rungs).
(b) Comparison of different averaging methods of the MLE curves. In the background, faint lines show every 5th curve out of the 60 MLEs to highlight where the averaged curves fall relative to all repetitions.
}
\label{fig:MLE}
\end{figure}

Fig.~\ref{fig:MLE}(a) shows all 60 fitted curves. There is a clear clustering of the curves, indicating that repetitions of the experiment are largely identical, with very few outliers. 

In addition, we compare different averaging methods. Figure~\ref{fig:MLE}(b) shows that all averaging methods yield very similar curves. The mean corresponds to a simple arithmetic average. The median picks the middle of the sorted values and does not assume a Gaussian distribution across the 60 repetitions. Finally, the pooled-data approach applies MLE under the assumption that the 60 curves are limited by shot noise, performing a single multinomial MLE with an effective $300\times60$ shots per point, so in this case the optimization finds the underlying probability curves that are most consistent with the experimental data per scan point and across all repetitions combined. The close agreement between the pooled MLE and the average of the 60 individually fitted curves indicates that the remaining noise is dominated by shot noise. Although the pooled method is computationally less demanding, fitting 60 curves allows a more unbiased estimation of the true values of triplet number and variance and simplifies the estimation of the error bars. In particular, implementing the SPAM correction fixes most of the systematic error, so the exact optimizer does not matter much. Therefore, in the main text we report the mean values of the 60 MLE curves. We also choose the mean of the 60 repetitions to capture possible experimental drift, which would be obscured in the pooled approach.

Since the experiment is repeated 60 times, this effectively provides a bootstrap-like sampling of the estimator. In fact it provides an empirical distribution across repetitions. The error bars are taken as $\sigma$, the standard deviation across the 60 reconstructed curves. This captures both shot noise and slow drift.

As mentioned above, the same procedure applies to the measurements of the phase diagram and the correlation plots. The only difference is in the measurement statistics. For the phase diagram, we take 200 shots per pixel and repeat the experiment three times. For the correlation functions, we perform a more precise measurement: for each value of $\overline{J_{||}}$ we measure an individual SPAM matrix, and the correlation measurements are taken with 200 shots per $x$ point and 40 repetitions of the experiment.
\clearpage

\subsection{Comparison to Gaussian MLE}
\label{sec:Gaussian}

\begin{figure}[h]
\centering
\includegraphics[width=\textwidth]{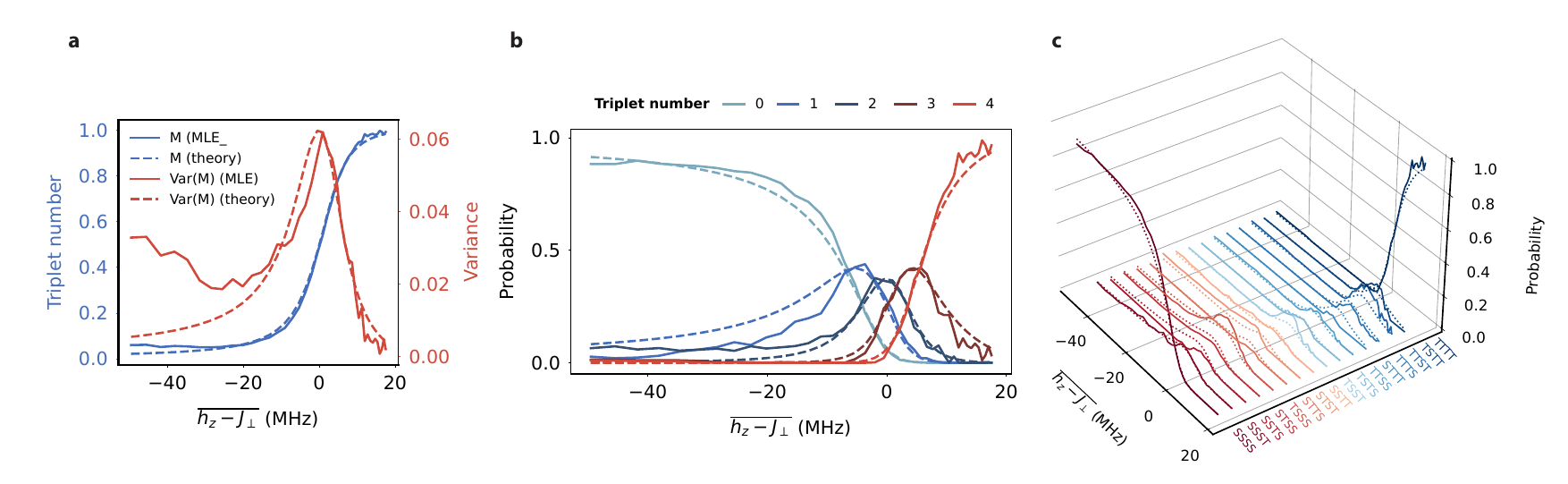}
\caption{%
\textbf{Data processing with Gaussian assumption.}
All measurements are identical to those in Fig.~\ref{fig:fig2} but the data is processed differently: instead of fitting each experimental repetition $r$ independently and averaging the results, all repetitions are pooled, and we process the entire dataset per point as a single measurement. (a) Triplet number (per site) and triplet number variance.
(b) Probabilities for obtaining $N$ triplet over all sites.
(c) Probability distribution over all possible outcomes -- combinations of $S$ and $T$ states -- over four sites.
}
\label{fig:gaussian}
\end{figure}

Under certain assumptions, the multinomial MLE procedure described above can be approximated by a Gaussian distribution. In particular, if the sample size is large and we treat the 60 repetitions of 300 shots per scan point as one pooled dataset, then each scan point effectively contains
\[
N = 300 \times 60 = 18{,}000
\]
shots. This approximation assumes that there is no significant drift between repetitions, so that all repetitions at the same \(x\) are drawn from the same underlying probability distribution.

For a fixed scan point \(x\), let
\[
C^{(i)}_{x}
=
\sum_{r=1}^{60} c^{(i)}_{r,x}
\]
be the pooled number of observed outcomes \(i\), where \(i=1,\dots,16\) and $\sum_{i=1}^{16} C^{(i)}_{x}=N$.

Data points have the form of a Gaussian distribution centered at
\[
C^{(i)}_{x}=N q^{(i)}_{x}.
\]
Thus, in the large-sample limit, the multinomial distribution over the measured counts can be approximated by a Gaussian distribution around the predicted measured probabilities \(q_{x}=Mp_{x}\).

In this approximation, instead of minimizing the multinomial negative log-likelihood
\[
-
\sum_{x,i}
C^{(i)}_{x}
\log\!\big((M p_{x})_i\big),
\]
one may use the Gaussian quadratic form
\[
\frac{1}{2}
\sum_{x,i}
\frac{
\left[
C^{(i)}_{x}
-
N(Mp_{x})_i
\right]^2
}{
N(Mp_{x})_i
}.
\]

Therefore, when the repetitions are pooled and the effective number of shots per scan point is large, the multinomial MLE problem can be viewed as a Gaussian approximation to the same count statistics, with the Gaussian centered at the SPAM-corrected predicted counts
\[
N(Mp_{x})_i.
\]

Together with assuming shot noise, another limitation is that the Gaussian approximation does not naturally enforce the physical constraints on the probability distribution ($p^{(i)}_{x}\geq 0$ and 
$\sum_{i=1}^{16}p^{(i)}_{x}=1$). Therefore we explicitly imposed these constraints via bounded optimization when producing the plots of Fig.~\ref{fig:gaussian}.

\clearpage

\subsection{Calibration of the rung exchange}
\label{sec:Jperp}

\begin{figure}[b]
\centering
\includegraphics[width=\textwidth]{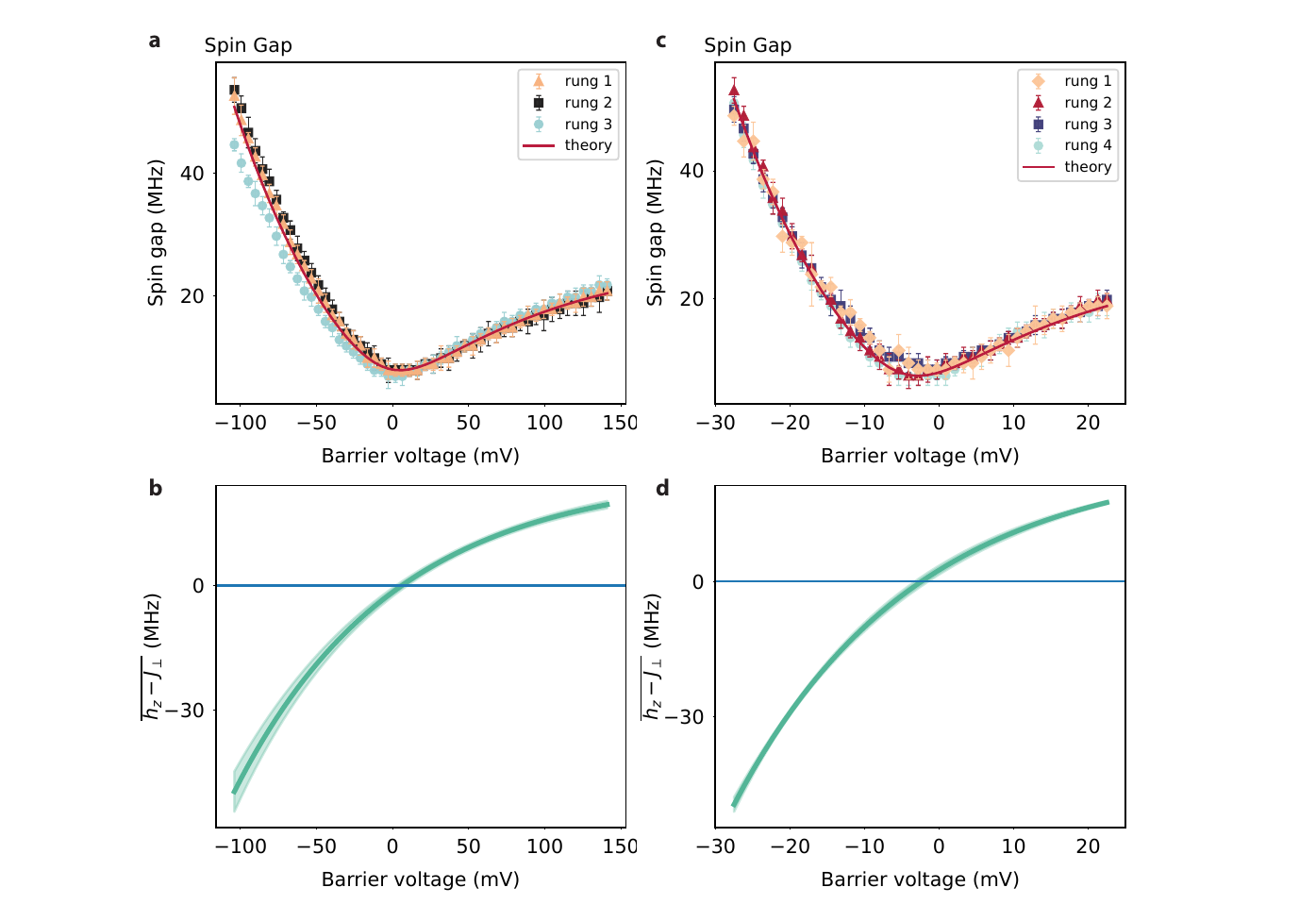}
\caption{%
\textbf{Calibration of rung-coupling sweeps.}
(a) Three-rung data, corresponding to Fig.~\ref{fig:fig3} and Fig.  ~\ref{fig:fig4}.
All three rungs are pulsed simultaneously as follows: a singlet is prepared on each rung, after which a $30$~ns pulse is applied to all rungs, implementing a $\pi/2$ rotation on the Bloch sphere.
The system then evolves for a waiting time of up to $1~\mu$s.
A second $\pi/2$ rotation is subsequently applied to project the state back onto the $ST_{-}$ basis, after which all rungs are measured, following a pulse sequence similar to that shown in Fig.~\ref{fig:fig3}.
The Fourier transforms of the resulting oscillations are shown in this panel, with each curve corresponding to a different rung.
The theoretical curve is obtained by solving the Hamiltonian in Eq.~(1).
The $x$-axis corresponds to the sweep of rung~3; however, each rung is swept over its own individual voltage range.
Specifically, the sweep ranges for rungs~1,~2, and~3 are $[-20,\,26]$, $[-116,\,67]$, and $[-104,\,141]$~mV, respectively.
The differences in these ranges arise from different lever arms of the corresponding gates $b_{15}$, $b_{26}$, and $b_{37}$.
Error bars represent the half-width at half-maximum (HWHM) of the FFT peak at each gate voltage point.
(b) Each curve in (a) is fitted with
$\hbar f_i = \sqrt{(h_{z,i} - J_{\perp,i})^2 + \Delta_{ST_{-},i}^2}$,
from which we extract $\overline{h_z - J_{\perp}}$, defined as the average over the three curves. The shaded region represents the standard deviation across qubits, highlighting the experimental variability among them.
(c) Analogous data for four rungs, corresponding to Fig.~\ref{fig:fig2}.
Here, the $x$-axis corresponds to the sweep of rung~4; as in the three-rung case, each rung is swept over its own individual voltage range.
The sweep ranges for rungs~1,~2,~3, and~4 are
$[-30,\,22.5]$, $[-65,\,123]$, $[-135,\,115]$, and $[-27.5,\,22.5]$~mV, respectively.
The difference from the three-rung case originates from a different overall device configuration and tuning.
(d) Same as (b), but for four rungs.
}
\label{fig:Jperp}
\end{figure}

In semiconductor heterostructures, both the $g$-factor and gate lever arms vary across the device~\cite{Zhang2025UniversalControlST, SweetSpotGeHoleSpinQubit}. In particular, because the gates are fabricated in different layers, the voltage amplitudes required to achieve a given exchange coupling differ between rung dimers. To mitigate these effects, we tune the voltage ranges such that, during a sweep of $J_{\perp}$, all rung dimers are driven through the same relative ratio of exchange and Zeeman energies.

To calibrate the $J_{\perp}$ sweep for each dimer, we perform simultaneous operations in the following manner. Using the language of $S-T_-$ qubits, the qubits in all dimers are initialized in the singlet state $S$ deep in the $(0,2)/(2,0)$ charge configuration, followed by a diabatic ramp to the respective centers of the $(1,1)$ charge regions. Subsequently, the qubits in all dimers are simultaneously rotated into the $xy$-plane of the qubit Bloch sphere with a $\pi/2$  ($30~\mathrm{ns}$) pulse. Oscillations about the $z$-axis occur at frequencies given by $(h_{z,i} - J_{\perp,i})$, and are measured for evolution times up to $1~\mu\mathrm{s}$. After subsequent additional $\pi/2$ pulses,  the dimers can be read out in the singlet-triplet basis.

After performing a Fourier analysis of the oscillations and extracting the corresponding frequencies, we observe that they are very close in value between the three rungs, indicating that the chosen pulse ranges are appropriate (Fig.~\ref{fig:Jperp}(a,c)). The frequency of the oscillations for each dimer is fitted with
\begin{equation}
h  f_i = \sqrt{(h_{z,i} - J_{\perp,i})^2 + \Delta^2_{ST_{-},i}},
\end{equation}
where $\Delta_{ST_{-},i}$ is the intrinsic spin-orbit coupling energy. We tune the barriers such that all dimers pass through the point where $h_{z,i} = J_{\perp,i}$ at approximately 0 mV; the frequency at this point is determined solely by $\Delta_{ST_{-},i}$.

Using the individually extracted parameters for each dimer, we compute $(h_{z,i} - J_{\perp,i})$. The average across three (four) rungs is shown in Fig.~\ref{fig:Jperp}(b,d). This averaged quantity is used as the $x$-axis in  Figs.~\ref{fig:fig2},~\ref{fig:fig3}, and~\ref{fig:fig4} of the main text. The measurements in Fig.~\ref{fig:fig4} were performed under slightly different detuning conditions; however, the voltage spans cover exactly the same range, since the position of $h_{z,i} \approx J_{\perp,i}$ was calibrated beforehand, and the voltage pulse limits were rescaled accordingly.

\clearpage

\subsection{$J_{\parallel}$ calibration}

\label{sec:Jpar}

The calibration procedure for the horizontal couplings is performed by tracking the $ST_{0}$ oscillation frequency as a function of magnetic field, and determining the minimum of the curve
\begin{equation} 
\label{eq:fit_J_parallel}
h f(B) = \sqrt{(\mu_{B} \Delta g_{ij}(B-B_0))^2 + J_{\parallel}^2},
\end{equation}
where $\Delta g_{ij}$ is the $g$-factor difference between neighboring dots, $\mu_B$ is Bohr magneton, and $B_0$ accounts for an offset field. During calibration of $J_\parallel$ for each pair, the surrounding dots are empty.
The minimum of the curve corresponds to the exchange interaction
$J_{\parallel}$ \cite{jirovec2021singletqubit}. 

\subsubsection{$J_{\parallel}$ calibration for four dimers: low $J_{\parallel}$ limit}

\begin{figure}[b]
\centering
\includegraphics[width=\textwidth]{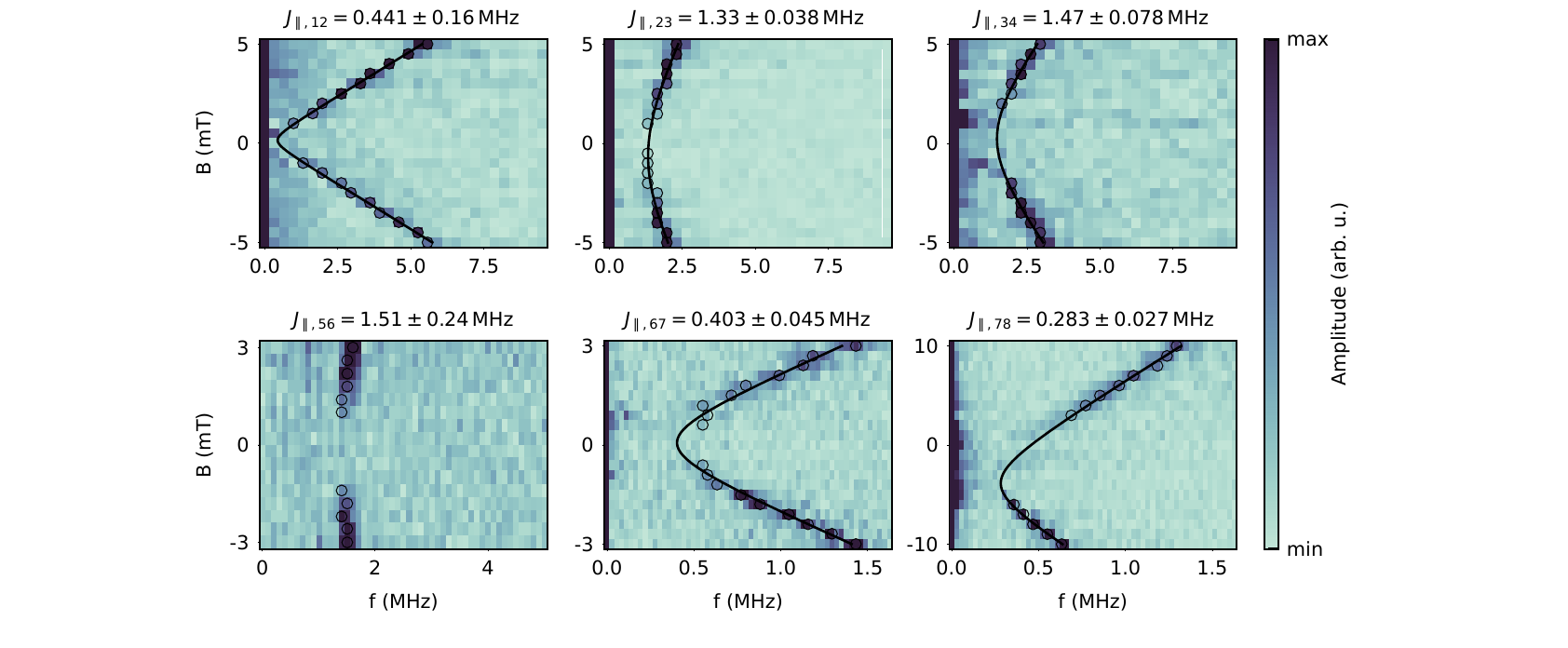}
\caption{%
\textbf{Calibration of inter-rung couplings for four dimers in the low-coupling limit.}
Panels correspond to the device layout of the barriers connecting the respective dots $i$ and $j$, which control the corresponding exchange coupling $J_{\parallel,ij}$. The values of the exchange interaction are obtained from the fits shown in each panel. The error bar in the panel corresponding to $J_{\parallel,56}$ represents the Gaussian standard deviation of the peak width, while for all other panels the uncertainty is estimated using a bootstrap method. Color scale represents FFT amplitude (triplet probability for $J_{\parallel, 56}$), normalized per panel. The first frequency bin is excluded from the normalization to suppress low-frequency components.
}
\label{fig:Jpar_4Q}
\end{figure}

For the case of four dimers, we perform the calibration described above for all relevant barriers connecting dots along the legs. Fourier transforms of the corresponding oscillations as a function of the external magnetic field are shown in Fig.~\ref{fig:Jpar_4Q}. For calibration of $J_{\parallel,56}$, we used a different procedure, directly performing microwave spectroscopy by applying a microwave signal while sweeping the external magnetic field. In the corresponding panel, the signal is shown after thresholding; no Fourier transform is required.

Each dataset is then processed in the following manner. Starting from the edges of the plot, the maximum of the FFT amplitude is identified (for pair 56, the maximum of the measured signal amplitude is used directly). The algorithm then proceeds toward the center of the plot, searching for the next peak within a frequency window of $1.5~\mathrm{MHz}$. Low-frequency points below $0.5~\mathrm{MHz}$ are excluded from the analysis, as well as points near zero magnetic field. Specifically, masks of $\pm 0.6~\mathrm{mT}$ for pair 56, $\pm 2~\mathrm{mT}$ for pair 34, $+2/-5~\mathrm{mT}$ for pair 78, and $\pm 0.5~\mathrm{mT}$ for all other pairs are applied. These exclusions prevent the selection of outliers when the FFT amplitude decreases near the center and/or becomes affected by the $ST_{-}$ component. For pair 78, the offset field $B_0$ is larger than for the other pairs, which motivates the asymmetric masking window.

The selected points are then fitted using Eq.~\ref{eq:fit_J_parallel}; the resulting fits are shown as black lines in Fig.~\ref{fig:Jpar_4Q}. Pair 56 is not fitted, as it exhibits only a weak dependence on the external magnetic field. Therefore, $J_{\parallel}$ is estimated directly from the measured value at $2~\mathrm{mT}$. In all other cases, the minimum of the fitted curve is taken as an estimate of the corresponding $J_{\parallel}$.

\subsubsection{$J_{\texorpdfstring{_{\parallel}}{||}}$ calibration for 3 dimers: sweeping $J_{\texorpdfstring{_{\parallel}}{||}}$}
\subsubsecmultitrue

\begin{figure}[b]
\centering
\includegraphics[width=\textwidth]{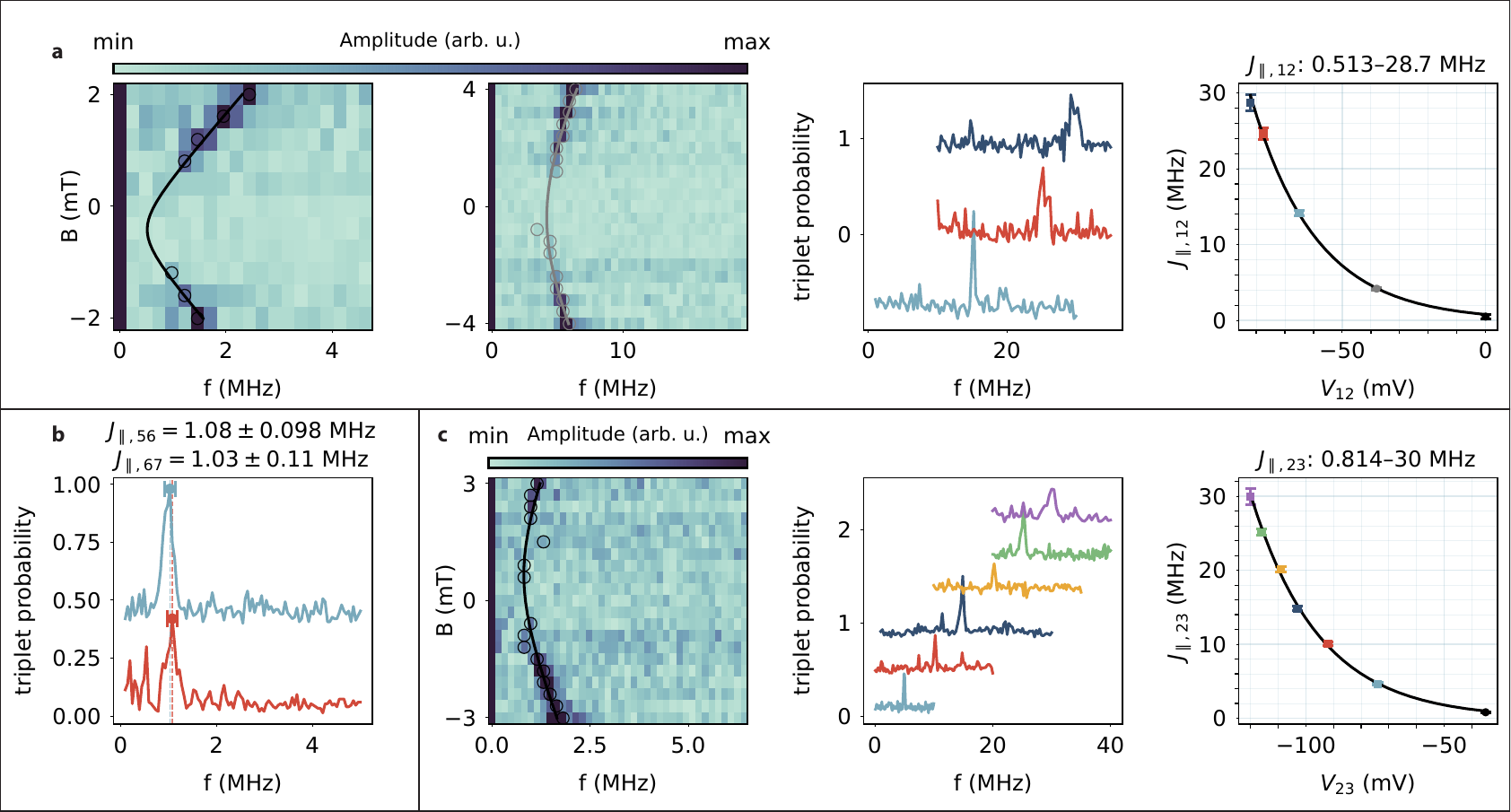}
\caption{
\textbf{Calibration of inter-rung couplings for three dimers.}
(a) Calibration scans used for extraction of $J_{\parallel, 12}$;
the first two panels follow the procedure of Fig.~\ref{fig:Jpar_4Q} and are processed in the same way, with voltage pulses of $0$ mV (reaching $J_{\parallel,12} = 0.513 \pm 0.28$ MHz) and $-38$ mV ($J_{\parallel,12} = 4.19 \pm 0.19$ MHz), respectively.
The third panel shows 1D scans of the $ST_0$ oscillation frequency for three values of the voltage pulses at a fixed field of 5 mT, whereby we use $\overline{\Delta g_{12}}$ extracted from the panels on the left (we assume here that the $g$-factor difference does not vary with gate voltage).
All extracted $J_\parallel$ values are shown in the last panel and fitted with an exponential.
Error bars for the 2D FFT data are extracted using a bootstrapping method,
while for the 1D scans they reflect the Gaussian standard deviation inferred from the peak width.
(b) Extracted $ST_0$ oscillation frequencies for the lower leg, providing an upper bound on the corresponding $J_\parallel$. These are the lower leg couplings used during the acquisition of the phase diagram.
(c) Analogous panels as in (a) but for $J_{\parallel, 23}$.}
\label{fig:Jpar_3Q}
\end{figure}

Calibration of the horizontal-coupling sweeps is carried out via a series of sequential measurements of the $ST_0$ oscillation frequency, either as a function of the external magnetic field or at a fixed magnetic field $B = 5~\mathrm{mT}$. The resulting data acquired while sweeping the magnetic field are fitted with Eq.~\ref{eq:fit_J_parallel} to extract the values of $J_{\parallel}$ for each applied voltage pulse amplitude. From the same fits we also extract $\overline{\Delta g_{12}}$ and $\overline{\Delta g_{23}}$.  Using the same expression, the bare exchange is obtained from 1D frequency scans at a fixed magnetic field.

The resulting curves for the upper leg are then fitted with an exponential function of $J_\parallel$ versus the corresponding virtual gate voltage. The obtained ranges for $J_\parallel$ are $0.513$–$28.7~\mathrm{MHz}$ and $0.814$–$30~\mathrm{MHz}$ for $J_{\parallel,12}$ and $J_{\parallel,23}$, respectively. Figure~\ref{fig:Jpar_3Q} (a, c) shows the corresponding calibration sweeps together with exponential fits. For the FFT data sets, the data processing is identical to that described in the previous subsection.

Figure~\ref{fig:Jpar_3Q} (b) shows the frequency of the $ST_0$ oscillation for pairs 56 and 67 at a magnetic field of 5 mT. For simplicity, we do not show an explicit calibration for these pairs. Since the measurements are performed at finite magnetic field, the extracted frequencies provide an upper estimate of the corresponding couplings $J_{\parallel,56}$ and $J_{\parallel,67}$.

\begin{figure}[t]
\centering
\includegraphics[width=\textwidth]{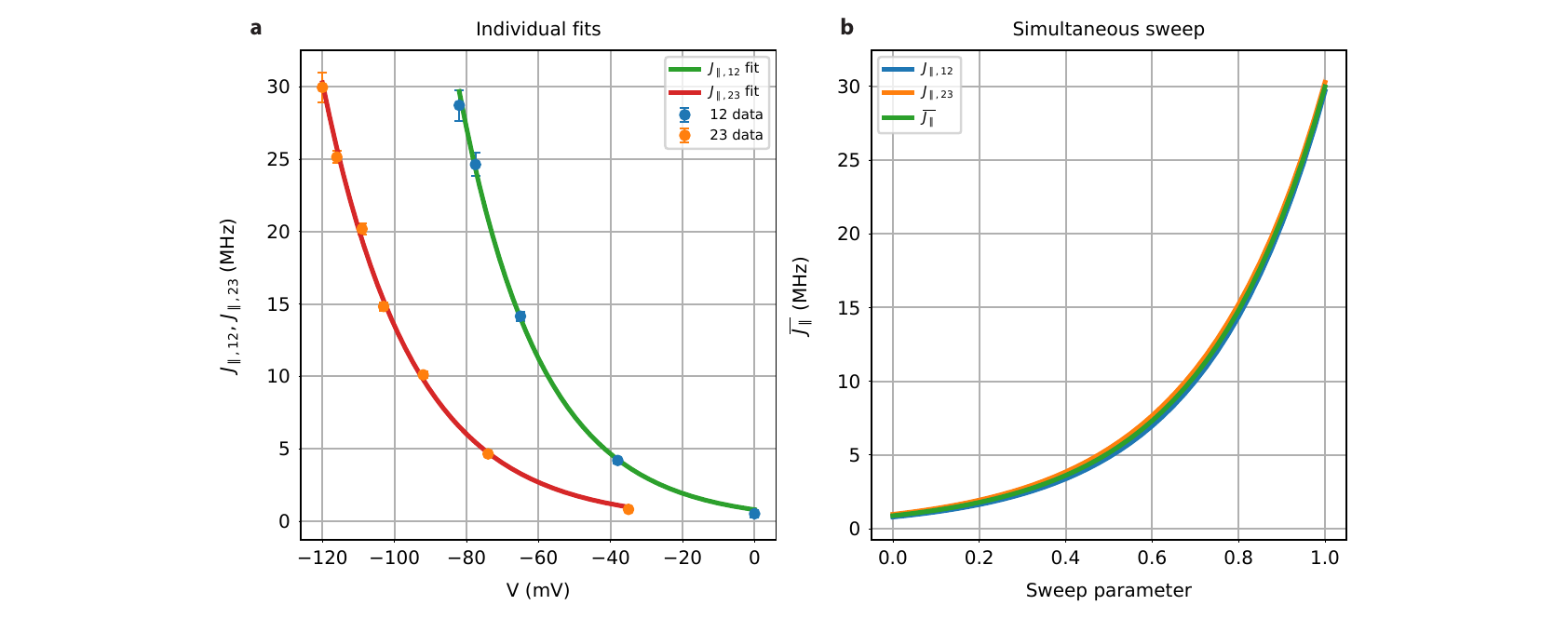}
\caption{%
\textbf{Universal $\overline{J_{\parallel}}$ axis for the phase diagram.}
(a) $J_{12}$ and $J_{23}$ as a function of the barrier voltage. Scatter points extracted from the data in Fig.~\ref{fig:Jpar_3Q} are fitted with an exponential dependence, relating exchange values to barrier voltages. This is shown to compare the relative ranges of reachable exchange values. 
(b) The same sweeps of the leg coupling exchange as a function of an arbitrary sweeping parameter, chosen such that the curves are nearly overlaid with one another. The average of these two curves is used in the main text as $J_{\parallel}$.
}
\label{fig:sim_sweep}
\end{figure}

To produce a universal axis for $\overline{J_{\parallel}}$ used in the phase diagram, we directly compare the ranges spanned by $J_{\parallel,12}$ and $J_{\parallel,23}$ and take the average between the corresponding exponential fits to construct a single calibration axis. This procedure is illustrated in Fig.~\ref{fig:sim_sweep}. We note that the averaging combines nearly identical curves, as the voltage ranges are carefully chosen to yield similar $J_{\parallel}$ values, and the barriers between dimers~1,~2, and~3 have comparable lever arms and sweep ranges. Throughout this paper, when we use the notation $J_{\parallel}$, it is an average value, shown in green in ~\ref{fig:sim_sweep} (b).

\clearpage

\subsection{Theoretical model}
\label{sec:theory}

\subsubsection{Hamiltonian}
\label{sec:hamiltonian}

We write down the Hamiltonian describing a two-leg quantum dot ladder with spinful electrons (or holes), incorporating hopping, spin--orbit interaction (SOI), on-site interactions, and a Zeeman field:

\begin{align}
H &=
- t_\perp 
\sum^L_{\substack{i=1 \\ \sigma=\uparrow,\downarrow}}
\Big( c_{t,i,\sigma}^\dagger c_{b,i,\sigma} + \mathrm{h.c.}\Big)
- t_{\parallel}
\sum^{L-1}_{\substack{\ell=t,b \\ i=1 \\ \sigma=\uparrow,\downarrow}}
\Big( c_{\ell,i,\sigma}^\dagger c_{\ell,i+1,\sigma} + \mathrm{h.c.}\Big)
- i t_{\perp}^{\mathrm{SO}} 
\sum^L_{\substack{i=1 \\ \alpha,\beta=\uparrow,\downarrow}}
\Big( c_{t,i,\alpha}^\dagger \sigma^x_{\alpha\beta} c_{b,i,\beta} + \mathrm{h.c.}\Big)
\nonumber\\
&\quad - i t_{\parallel}^{\mathrm{SO}} 
\sum^L_{\substack{\ell=t,b \\ i=1 \\ \sigma=\uparrow,\downarrow}}
\Big( c_{l,i,\alpha}^\dagger \sigma^z_{\alpha\beta} c_{l,i+1,\beta} + \mathrm{h.c.}\Big)
+ U 
\sum^L_{\substack{\ell=t,b \\ i=1}}
n_{\ell,i,\uparrow} n_{\ell,i,\downarrow}
+ \epsilon \sum^L_{i=1} \big(n_{t,i}-n_{b,i}\big)
+ \mu_B B 
\sum^L_{\substack{\ell=t,b \\ i=1}}
g_{\ell,i}\, S^z_{\ell,i}.
\label{eq:fermion_ham_tb}
\end{align}

The first and second terms represent spin-conserving tunneling between and within the two legs of the ladder (labeled by $\ell=t,b$) at each site $i$, with amplitude $t_\perp$ controlling the coupling across the rungs and $t_\parallel$ the coupling between adjacent sites within each leg. For the sake of simiplicity, we take the leg coupling $t_\parallel$ to be the same on both the legs. However, the derivation can be readily generalized to the case where the two legs have different leg couplings.

The third and fourth terms describe spin--orbit-coupled hopping. 
Microscopically, SOI in a crystal can be written as $H_{\mathrm{SOI}} \propto \boldsymbol{\sigma} \cdot (\mathbf{p} \times \mathbf{E})$, where $\mathbf{p}$ is the electron momentum and $\mathbf{E}$ is a static electric field breaking inversion symmetry. In the ladder geometry, momentum along the rungs is $\mathbf{p} = p_z \hat{\mathbf{z}}$, while the inversion-breaking field points along $-\hat{\mathbf{y}}$ (e.g., from a substrate or gating). 
Thus, the SOI couples motion along the rungs (legs) to spin along $\hat{\mathbf{x}}$ ($\hat{\mathbf{z}}$), producing spin-flip hopping between the two legs (adjacent rungs) in the $\{\uparrow_z,\downarrow_z\}$ basis. The coupling strength $t_\perp^{\mathrm{SO}}$ ($t_\parallel^{\mathrm{SO}}$) sets the amplitude of this process. Note that we find $t_\parallel^{\mathrm{SO}} \ll t_\perp^{\mathrm{SO}}$.

The fifth term accounts for the on-site Hubbard repulsion $U$ between opposite spins on the same dot. 
The sixth term represents an energy offset $\epsilon$ between the two legs. 
The final term describes Zeeman coupling to an external magnetic field $B$, where $\mu_B$ is the Bohr magneton and $g_{\ell, i}$ is the effective site-dependent $g$-factor. Since the applied magnetic field ($B \approx 5\,\mathrm{mT}$) is predominantly in-plane, the flux threading a lattice plaquette is negligible relative to the flux quantum. Accordingly, Peierls phases in the tunneling amplitudes are ignored.

In the strongly interacting regime, where $U$ is much larger than the tunneling amplitudes (e.g., $U\sim300~\mathrm{GHz}$ and $t_{\perp},t_{\parallel}\sim2~\mathrm{GHz}$), double occupancy is energetically suppressed. 
It is therefore appropriate to project the fermionic Hamiltonian into the singly occupied subspace. 
Suppressing orbital indices and temporarily omitting detuning and Zeeman terms, Eq.~ (1) can be written as
\begin{equation}
H = \sum_{ij} \vec{c}_i^\dagger h_{ij} \vec{c}_j
+ U \sum_i n_{i\uparrow} n_{i\downarrow},
\end{equation}
with
\begin{equation}
\vec{c}_i^\dagger = \left(c_{i\uparrow}^\dagger, c_{i\downarrow}^\dagger \right),
\qquad
h_{ij} = t_{ij}\mathbb{I} + i \vec{v}_{ij}\cdot\vec{\sigma}.
\end{equation}

Using Rayleigh--Schr\"odinger perturbation theory at half filling, the second-order correction to the effective Hamiltonian is
\begin{equation}
H_{\mathrm{eff}}^{(2)}
\approx
- \sum_{\alpha\beta}\sum_{n\neq\alpha\beta}
\left[
\frac{\langle \alpha | \hat T^\dagger | n \rangle \langle n | \hat T | \beta \rangle}{E_n-E_0}
+
\frac{\langle \alpha | \hat T | n \rangle \langle n | \hat T^\dagger | \beta \rangle}{E_n-E_0}
\right]
|\alpha\rangle\langle\beta|,
\end{equation}
where $\hat T=\vec c_i^\dagger h_{ij}\vec c_j$ is the kinetic operator and the indices $\alpha$ and $\beta$ run over the single particle manifold and the index $n$ is over the doubly occupied subspace. Using $\langle \alpha|\hat T|\beta\rangle=0$, completeness of singly and doubly occupied subspaces, and $E_n-E_0=U$, the calculation reduces to evaluating $T^2$. 
With the Abrikosov fermion representation
\begin{align}
\mathbf S_i &= \frac{1}{2}\,\mathbf c_i^\dagger \boldsymbol{\sigma}\,\mathbf c_i,
\\
\mathbf c_i \mathbf c_i^\dagger &= \frac{1}{2}\mathbb I - \mathbf S_i\cdot\boldsymbol{\sigma},
\end{align}
one obtains an effective spin Hamiltonian
\begin{align}
H &= 
J_{\perp}\sum^L_i \vec S_{t, i}\cdot\vec S_{b, i}
+ J_{\parallel}\sum^{L-1}_{i,\ell}\vec S_{\ell, i}\cdot\vec S_{\ell, i+1}
+ \vec D_{\perp}\cdot\sum_i^L (\vec S_{t, i}\times\vec S_{b, i})
+ \vec D_{\parallel}\cdot\sum^{L-1}_{\ell, i} (\vec S_{\ell,i}\times\vec S_{\ell, i+1})
\nonumber\\
&\quad
+ \sum^L_i (\vec\gamma_{\perp}\cdot\vec S_{t, i})(\vec\gamma_{\perp}\cdot\vec S_{b, i})
+ \sum^{L-1}_{\ell, i} (\vec\gamma_{\parallel}\cdot\vec S_{\ell, i})(\vec\gamma_{\parallel}\cdot\vec S_{\ell, i+1})
+ \mu_B \vec B \cdot \sum_{\ell, i} g_{\ell, i}\vec S_{\ell, i}.
\label{Eq_spinhamfull}
\end{align}

To connect with the experiment, we restore orbital indices, detuning and  and impose the specific directions of spin-dependent hopping corresponding to an out-of-plane electric field.
This yields the component form
\begin{align}
H &= 
J_{\perp}\sum^L_i \vec S_{t,i}\cdot\vec S_{b, i}
+ J_{\parallel}\sum_{\ell, i}\vec S_{i,\ell}\cdot\vec S_{\ell, i+1}
+ D_{\perp}^x \sum^{L-1}_i \left(S_{t, i}^y S_{b, i}^z - S_{t, i}^z S_{b, i}^y\right)
+ D_{\parallel}^z \sum^{L-1}_{\ell, i}
\left(S_{\ell, i}^y S_{\ell, i+1}^x - S_{\ell, i}^x S_{\ell, i+1}^y\right)
\nonumber\\
&\quad
+ (\gamma_{\perp}^x)^2 \sum^L_i S_{t, i}^x S_{b, i}^x
+ (\gamma_{\parallel}^z)^2 \sum^{L-1}_{\ell, i} S_{\ell, i}^z S_{\ell, i+1}^z
+ \mu_B B^z \sum^L_{\ell, i} g_{\ell, i} S^z_{\ell, i}.
\label{Eq_spin_ham}
\end{align}

The coupling constants in terms of the fermionic parameters are
\begin{align}
J_{\perp} &= \frac{4U}{U^2-\epsilon^2}\left(t_{\perp}^2 - t_{\perp, SO}^2\right),
&
J_{\parallel} &= \frac{4U}{U^2-\epsilon^2}\left(t_{\parallel}^2 - t_{\parallel, SO}^2\right),
\\
D_{\perp}^x &= \frac{8U}{U^2-\epsilon^2}\left(t_{\perp}\,t_{\perp, SO}\right),
&
D_{\parallel}^z &= \frac{8U}{U^2-\epsilon^2}\left(t_{\parallel}\,t_{\parallel, SO}\right),
\\
(\gamma_{\perp}^x)^2 &= \frac{8U}{U^2-\epsilon^2} t_{\perp, SO}^2,
&
(\gamma_{\parallel}^z)^2 &= \frac{8U}{U^2-\epsilon^2} t_{\parallel, SO}^2.
\label{Eq:paramrelations}
\end{align}

Here $t_{\perp}$ and $t_{\parallel}$ are the spin-conserving tunneling amplitudes across rungs and along legs, while $t_{\parallel SO}$ and $t_{\parallel SO}$ denote their spin-flipping counterparts. 
The denominators arise from virtual charge excitations with inter-leg detuning $\epsilon$, demonstrating how charge fluctuations mediate anisotropic spin interactions.

Given that we experimentally find very small $S-T_-$ anti-crossings for the pairs along the legs, indicating weak $t_{\parallel SO}$. In what follows, we neglect for the moment $t_{\parallel SO}$ to simplify the analysis. Therefore we arrive at the Hamiltonian

\begin{align}
H &=
J_{\perp} \sum_{i=1}^{L} \vec{S}_{t,i}\cdot\vec{S}_{b,i}
+ J_{\parallel}\sum_{\ell=t,b}\sum_{i=1}^{L-1}
\vec{S}_{\ell,i}\cdot\vec{S}_{\ell,i+1}
+ D_{\perp}^{x} \sum_{i=1}^{L}
\left(
S_{t,i}^y S_{b,i}^z - S_{t,i}^z S_{b,i}^y
\right)
\nonumber\\
&\quad
+ (\gamma_{\perp}^x)^2 \sum_{i=1}^{L}
S_{t,i}^x S_{b,i}^x
+ \mu_B B^z \sum_{\ell=t,b}\sum_{i=1}^{L}
g_{\ell,i} S_{\ell,i}^z \, .
\label{Eq:mainH1}
\end{align}

Here, performing calculations similar to those in Ref.~\cite{CoupledLaddersMagneticField}, we 
project the original Hamiltonian onto the singlet-triplet subspace:

\[
|\tilde{\uparrow}\rangle =
\frac{1}{\sqrt{2}}
\left[
|\uparrow \downarrow \rangle
-
|\downarrow \uparrow \rangle
\right]
\]

\begin{equation}
|\tilde{\downarrow}\rangle = |\downarrow \downarrow \rangle
\end{equation}

Thus, the singlet, i.e., the absence of a triplet excitation, corresponds to the new
state $|\tilde{\uparrow}\rangle$, while the presence of a triplet excitation,
in our case $T_{-}$, corresponds to $|\tilde{\downarrow}\rangle$. Other degrees of freedom are ignored as they are higher in energy.
% and not reachable with adiabatic state preparation.

Then,

\begin{equation}
S^{\pm}_1|\tilde{\downarrow}\rangle
=
 \frac{1}{\sqrt{2}}
\tilde S^{\pm}
|\tilde{\downarrow}\rangle, \quad
S^{\pm}_2|\tilde{\downarrow}\rangle
=
 -\frac{1}{\sqrt{2}}
\tilde S^{\pm}
|\tilde{\downarrow}\rangle \;,
\end{equation}

\begin{equation}
S^z_{1,2} |S_0 \rangle
=
0
=
-\frac{1}{4}
\left(
I - 2\tilde{S}^z
\right)
|\tilde{\uparrow} \rangle \;,
\end{equation}

\begin{equation}
S^z_{1,2} |T_{-} \rangle
= -\frac{1}{2} |\tilde{\downarrow} \rangle =
-\frac{1}{4}
\left(
I - 2\tilde{S}^z
\right)
|\tilde{\downarrow} \rangle \;.
\end{equation}

The Hamiltonian then becomes
\begin{equation}
\begin{aligned}
H &=
J_{\parallel} \sum_{i}^{L-1}
\left(
\tilde{S}_i^x \tilde{S}_{i+1}^x
+
\tilde{S}_i^y \tilde{S}_{i+1}^y
\right)
+
\frac{J_{\parallel}}{2}
\sum_{i}^{L-1}
\tilde{S}_i^z \tilde{S}_{i+1}^z
\\
&\quad+
\sum_{i}^{L}
\left(
h_{z, i}
-
J_{\perp}
-
\frac{J_{\parallel}}{2}
\right)
\tilde{S}_i^z
-\frac{D^x_\perp}{2\sqrt{2}}
\sum_{i}^{L}
\tilde{S}_i^y ,
\end{aligned}
\label{eqn:effspinchainham}
\end{equation}
which is the form that appears in the main text.

\subsubsection{Ground State Phases via Spiral Ansatz}
\label{sec:bethe}
Here we apply the variational principle to a spiral ansatz to analyze the ground-state phases in the absence and presence of the SOI term in the Hamiltonian in Eq.~\eqref{eqn:effspinchainham}. From the previous sections, assuming $D = D^x_{\perp}$ for simplicity, we obtain the SOI term (we here use site indices $j$ instead of $i$ to avoid confusion with the imaginary number)
\[
H_D =
-\frac{D}{2\sqrt{2}}
\sum_j \tilde{S}^y_j .
\]

Let
\[
\ket{\psi}
= \prod_{j}
\left[
\cos\left(\frac{\Theta}{2}\right)\ket{\tilde{\uparrow}}
+
e^{iQj}\sin\left(\frac{\Theta}{2}\right)\ket{\tilde{\downarrow}}
\right]_j.
\]

We also define uniform $\tilde{h} = h_z - J_{\perp} - \frac{J_{\parallel}}{2}$, ignoring small spatial variations in the g-factor, and rewrite the Hamiltonian in Eq.~\eqref{eqn:effspinchainham} in the simplified form
\begin{equation}
\begin{aligned}
H &=
J_{\parallel}\sum_{i}
\left(
\tilde{S}_i^x \tilde{S}_{i+1}^x
+
\tilde{S}_i^y \tilde{S}_{i+1}^y
+
\frac{1}{2}\tilde{S}_i^z \tilde{S}_{i+1}^z
\right)
+
\tilde{h}
\sum_{i}\tilde{S}_i^z
-\frac{D}{2\sqrt{2}}
\sum_{i}
\tilde{S}_i^y .
\end{aligned}
\end{equation}

Using the variational state above, we obtain
\[
\langle \tilde{S}^x_j\rangle
=
\frac{1}{2}
\langle \tilde{S}^{+}_j+\tilde{S}^{-}_j\rangle
=\frac{1}{2}\bra{\psi} 
\langle \tilde{S}^{+}_j+\tilde{S}^{-}_j\rangle \ket{\psi}
=\frac{1}{2}\sin\Theta\cos(Qj).
\]

Similarly,
\[
\langle \tilde{S}^y_j\rangle
=
\frac{1}{2}\sin\Theta\sin(Qj),
\]
and
\[
\langle \tilde{S}^z_j\rangle
=
\frac{1}{2}\cos\Theta .
\]

Therefore, for the expectation value of the Hamiltonian, we find
\[
\begin{aligned}
\langle H\rangle
&=
J_{\parallel}
\sum_j
\left[
\langle \tilde{S}^x_j\rangle
\langle \tilde{S}^x_{j+1}\rangle
+
\langle \tilde{S}^y_j\rangle
\langle \tilde{S}^y_{j+1}\rangle
+
\frac{1}{2}
\langle \tilde{S}^z_j\rangle
\langle \tilde{S}^z_{j+1}\rangle
\right]
+
\tilde{h}
\sum_j
\langle \tilde{S}^z_j\rangle
-
\frac{D}{2\sqrt{2}}
\sum_j
\langle \tilde{S}^y_j\rangle .
\end{aligned}
\]

Dropping the intermediate steps, and assuming periodic boundary conditions, we arrive at the following expectation value of the Hamiltonian:
\[
\langle H\rangle
=
\frac{J_{\parallel}L}{4}\sin^2\Theta \cos Q
+
\frac{J_{\parallel}L}{8}\cos^2\Theta
+
\frac{\tilde{h}L}{2}\cos\Theta
-
\frac{D}{4\sqrt{2}}\sin\Theta
\sum_j \sin(Qj).
\]

In the limit of no SOI, i.e. $D=0$, we need to satisfy
\[
\frac{\partial \langle H\rangle}{\partial Q}
=
-\frac{J_{\parallel}L}{4}\sin^2\Theta \sin Q
=
0
\]
and
\[
\frac{\partial^2 \langle H\rangle}{\partial Q^2}
=
-\frac{J_{\parallel}L}{4}\sin^2\Theta \cos Q
>
0 .
\]
This gives a stable solution at $Q=\pi$. Additionally, the values $\Theta=0$, $\Theta\neq 0,\pi$, and $\Theta=\pi$ correspond to the RS, CAFM, and FP phases, respectively. $Q=\pi$ is accepted as a stable solution indicating antiferromagnetic ordering with wave vector $\pi$ in the absence of SOI.

For finite SOI, i.e. $D \neq 0$, we obtain
\[
\frac{\partial \langle H\rangle}{\partial Q}
=
-\frac{J_{\parallel}L}{4}\sin^2\Theta \sin Q
-
\frac{D}{4\sqrt{2}}\sin\Theta \sum_j j\cos(Qj).
\]

Assuming that $D$ is small compared to $J_{\parallel}$, we expand around the antiferromagnetic wave vector,
\[
Q = \pi + q .
\]
Using
\[
\sum_j j\cos(Qj) \sim \frac{L}{2},
\]
the minimization condition becomes
\[
\frac{\partial \langle H\rangle}{\partial Q}
\sim
-\frac{J_{\parallel}L}{4}\sin^2\Theta \sin(\pi+q)
-
\frac{D}{4\sqrt{2}}\sin\Theta \frac{L}{2}
=0 .
\]
Since $\sin(\pi+q)=-\sin q$, this gives
\[
J_{\parallel}\sin\Theta \sin q
\sim
-\frac{D}{2\sqrt{2}} .
\]
For $\Theta\neq 0$, the shift of the wave vector is therefore
\[
\sin q
\sim q
\sim
-\frac{D}{2\sqrt{2}J_{\parallel}\sin\Theta}
\sim 
-\frac{D}{\sqrt{2}J_{\parallel}}.
\]

\subsubsection{Phase Boundaries}
\label{sec:slopestheory}
We now determine the critical lines that define the structure of the phase diagram. We focus on the case without SOI and, for simplicity, replace the second-to-last term of the Hamiltonian in Eq.~(20) by an effective longitudinal magnetic field $\tilde{h}_z$, related to the physical field via $\tilde{h}_z = h_{z} - J_{\perp} - J_{\parallel}/2$, where we have again ignored small spatial variations of $h_{z}$.

We begin by considering the rung-singlet configuration $|\tilde{\uparrow}\tilde{\uparrow}\tilde{\uparrow}\rangle$. Its energy, calculated from Eq.~(20) for $L$ rungs, is
\begin{equation}
E_{|\tilde{\uparrow}\tilde{\uparrow}\tilde{\uparrow}\rangle}
= \frac{1}{4}\times\frac{J_{\parallel}}{2}\times L + \frac{\tilde{h}_z}{2}L.
\end{equation}
Next, we consider a single delocalized spin flip (or triplon insertion) on top of this background. We focus on the $k=\pi$ mode as it minimizes the kinetic energy of the delocalized spin flip: the hopping term $2 (J_{\parallel}/2) \cos k$ is most negative at $k=\pi$, making this the lowest-energy single-excitation state and thus the relevant one for determining the phase boundary. This yelds,
\begin{equation}
|k=\pi\rangle = \sum_i c_i \, |\tilde{\uparrow} \ldots \tilde{\downarrow}_i \ldots \tilde{\uparrow}\rangle,
\end{equation}
where $c_i$ is superposition with right sign structure consistent with $k = \pi$-mode. This  represents the onset of the CAFM phase near this boundary, with energy
\begin{equation}
E_{|k=\pi\rangle}
= \underbrace{-2\cdot\frac{J_{\parallel}}{2}}_{\text{hopping } \tilde{S}^+\tilde{S}^-}
\;\underbrace{{}+\frac{J_{\parallel}}{2}\cdot\frac{1}{4}(L-2) - \frac{J_{\parallel}}{2}\cdot\frac{1}{4}\cdot 2}_{\tilde{S}^z\tilde{S}^z \text{ interaction}}
\;\underbrace{{}+\frac{\tilde{h}_z}{2}(L-2)}_{\text{Zeeman field}}.
\end{equation}
The transition between these two states marks the crossover from the RS phase to the CAFM phase. Equating the two energies gives
\begin{equation}
\tilde{h}_{z,c_1} = -\frac{3J_{\parallel}}{2}.
\end{equation}
Using $\tilde{h}_z = h_{z} - J_{\perp} - J_{\parallel}/2$, we obtain the relation between the $x$- and $y$-axes of the phase diagram,
\begin{equation}
h_{z} - J_{\perp} = -J_{\parallel},
\end{equation}
which determines the slope of the transition from the RS phase to the CAFM phase.

Similarly, the fully polarized configuration $|\tilde{\downarrow}\tilde{\downarrow}\tilde{\downarrow}\rangle$ has energy
\begin{equation}
E_{|\tilde{\downarrow}\tilde{\downarrow}\tilde{\downarrow}\rangle}
= \frac{1}{4}\times\frac{J_{\parallel}}{2}\times L - \frac{\tilde{h}_z}{2}L.
\end{equation}
Next, consider a single delocalized spin flip (or a singlon insertion) on top of this fully polarized background,
\begin{equation}
|k=\pi\rangle = \sum_i c_i \, |\tilde{\downarrow} \ldots \tilde{\uparrow}_i \ldots \tilde{\downarrow}\rangle,
\end{equation}
again representing the onset of the CAFM phase, now near the FP boundary, with energy
\begin{equation}
E_{|k=\pi\rangle}
= \underbrace{-2\cdot\frac{J_{\parallel}}{2}}_{\text{hopping } \tilde{S}^+\tilde{S}^-}
\;\underbrace{{}+\frac{J_{\parallel}}{2}\cdot\frac{1}{4}(L-2) - \frac{J_{\parallel}}{2}\cdot\frac{1}{4}\cdot 2}_{\tilde{S}^z\tilde{S}^z \text{ interaction}}
\;\underbrace{{}-\frac{\tilde{h}_z}{2}(L-2)}_{\text{Zeeman field}}.
\end{equation}
The transition between these states marks the crossover from the CAFM phase to the FP phase. Equating the energies gives
\begin{equation}
\tilde{h}_{z,c_2} = \frac{3J_{\parallel}}{2},
\end{equation}
which, using the same substitution, yields the phase boundary
\begin{equation}
h_{z} - J_{\perp} = 2J_{\parallel}.
\end{equation}
Therefore, the phase diagram expands to both sides from the point $h_z \approx J_{\perp}$ as a function of $J_{\parallel}$, and the slope of the CAFM--FP boundary is twice as steep and opposite in sign to that of the RS--CAFM boundary.

\subsubsection{Limit of $J_{\perp}$ approaching 0}
\label{sec:rightlimit}
Within the finite-size limit of our device, we analyze a regime that shows the FP phase terminates at a critical leg coupling $J_{\parallel,C}$, beyond which the CAFM phase extends to occupy the remaining parameter space. Let us assume the extreme limit on the $x$-axis, with $J_{\perp} \rightarrow 0$. We also assume that one of the leg couplings is zero, while the other (say, the upper leg) has a finite exchange strength $J_{\parallel}$. We examine two trial states. The first is a fully polarized triplet product state $\ket{\phi}$, corresponding to the FP phase. The second state $\ket{\psi}$ retains a fully polarized lower leg but has a delocalized spin flip on the upper leg due to the finite exchange. The two lowest eigenstates states for a system with three rungs are:

\begin{equation}
\begin{aligned}
\ket{\phi} &=
\left| \begin{array}{ccc}
\downarrow & \downarrow & \downarrow \\
\downarrow & \downarrow & \downarrow
\end{array} \right\rangle ,
\\[10pt]
\ket{\psi} &=
\frac{1}{\sqrt{6}}
\left(
-
\left| \begin{array}{ccc}
\uparrow & \downarrow & \downarrow \\
\downarrow & \downarrow & \downarrow
\end{array} \right\rangle
+
2
\left| \begin{array}{ccc}
\downarrow & \uparrow & \downarrow \\
\downarrow & \downarrow & \downarrow
\end{array} \right\rangle
-
\left| \begin{array}{ccc}
\downarrow & \downarrow & \uparrow \\
\downarrow & \downarrow & \downarrow
\end{array} \right\rangle
\right) .
\end{aligned}
\end{equation}

The energy of $\ket{\phi}$ consists of a Zeeman term and the exchange-energy contribution from the two nearest-neighbor pairs on the upper leg (each with parallel spins), yielding
\begin{equation}
E_\phi = -3h_z + \frac{J_{\parallel}}{2}.
\end{equation}

For the state $\ket{\psi}$ the energy is
\begin{equation}
E_\psi = -J_{\parallel} - 2h_z.
\end{equation}

Equating these energies gives the critical field strength at which the two states are degenerate:
\begin{equation}
h_z = \frac{3}{2} J_{\parallel}.
\end{equation}

$J_{\parallel, C} = \frac{2}{3}h_z$ then marks the critical value of the leg coupling above which the fully polarized state is no longer reachable. The state $\ket{\psi}$ has a triplet number of $2.5$, which is less than the maximum value of $3$ expected in the fully polarized limit. As a consequence, in a system with three rungs at $J_{\perp} \rightarrow 0$ the plateau at triplet number $3$ is no longer accessible, resulting in only two first-order transitions in the triplet number.
\clearpage

\subsection{Theoretical fit and comparison with the experimental data}
\label{sec:theoreticalfit}

\subsubsection{Rung Parameter Estimation}
The Hamiltonian in Eq.~\eqref{eq:fermion_ham_tb} is defined by the parameters $t_\perp, t_\perp^{\mathrm{SO}}, U, \epsilon, g_1, g_2$ (rung indices suppressed for brevity). Since $g_1$ and $g_2$ are determined experimentally, it remains to fix $t_\perp, t_\perp^{\mathrm{SO}}, U$, and $\epsilon$ theoretically.
Our strategy is to use the experimentally measured singlet--triplet gap as a function of barrier voltage (Fig.~\ref{fig:Jperp})
and fit the corresponding theoretical expression to this data. 
Since the relation between the barrier voltage and the tunneling amplitude is not known \emph{a priori}, 
we assume an exponential dependence of the form
\begin{equation}
\label{eq:t(V)}
t(V) = A\, e^{-\alpha (V - V_0)} + t_\infty.
\end{equation}
Using this parametrization, we perform a global fit to the data and extract the effective parameters for each pair of sites.  

To derive the singlet-triplet gap, we begin with the spin Hamiltonian in Eq.~\eqref{Eq:mainH1} for a rung of a quantum dot ladder, incorporating Heisenberg exchange ($J$), Dzyaloshinskii–Moriya (DM) interaction ($D = D^x_\perp$), spin-orbit coupling (SOC) term ($\gamma^2$ = $(\gamma^x_\perp)^2$), and Zeeman splitting due to an external magnetic field $B_z$:
\begin{align}
H &= J \vec{S}_0 \cdot \vec{S}_1 + D(S_0^y S_1^z - S_0^z S_1^y) \notag \\
&+ \gamma^2 S_0^x S_1^x + \mu_B B_z (g_0 S_0^z + g_1 S_1^z)
\end{align}
where the relation of the parameters in this Hamiltonian to the bare parameters $t_\perp, t_\parallel, t_\perp^{\mathrm{SO}}, U, \epsilon, g_1, g_2$  is specified in Eq.~\eqref{Eq:paramrelations}.

\noindent We use the basis $\{S_0, T_-, T_+, T_0\}$ where
 $S_0$ is the singlet state $\frac{1}{\sqrt{2}}(\uparrow\downarrow - \downarrow\uparrow)$,
 $T_0$ is the triplet with zero spin projection $\frac{1}{\sqrt{2}}(\uparrow\downarrow + \downarrow\uparrow)$, $T_+ = |{\uparrow\uparrow}\rangle$ and $T_- = |{\downarrow\downarrow}\rangle$. The matrix representation of $H$ in this basis, defining $\Delta g = g_0 - g_1$, $\bar{g} = \frac{g_0 + g_1}{2}$ for brevity, is:

\begin{table}[h]
\centering
\begin{tabular}{c|cccc}
 & $S_0$ & $T_-$ & $T_+$ & $T_0$ \\
\hline
$S_0$ & $-\frac{3J}{4} - \frac{\gamma^2}{4}$ & $i D \frac{\sqrt{2}}{4}$ & $-i D \frac{\sqrt{2}}{4}$ & $\mu_B B_z \frac{\Delta g}{2}$ \\
$T_-$ & $-i D \frac{\sqrt{2}}{4}$ & $\frac{J}{4} - \mu_B B_z\bar{g}$ & $\frac{\gamma^2}{4}$ & 0 \\
$T_+$ & $i D \frac{\sqrt{2}}{4}$ & $\frac{\gamma^2}{4}$ & $\frac{J}{4} + \mu_B B_z \bar{g}$ & 0 \\
$T_0$ & $\mu_B B_z \frac{\Delta g}{2}$ & 0 & 0 & $\frac{J}{4} + \frac{\gamma^2}{4}$ \\
\end{tabular}
\end{table}

\noindent
We now derive an effective $2 \times 2$ Hamiltonian for the low-energy subspace (qubit). Define the projectors
\[
P = |S_0\rangle\langle S_0| + |T_l\rangle\langle T_l|, \quad Q = \mathbf{1} - P .
\]
Here $T_l$ is the lower-energy combination of $T_+$ and $T_-$, and the remaining states ($T_h$, $T_0$) form the high-energy subspace.

\noindent
Since $P+Q=\mathbf{1}$, projecting the Schr\"odinger equation $H|\psi\rangle=E|\psi\rangle$
with $P$ and $Q$ and inserting $\mathbf{1}=P+Q$ between $H$ and $|\psi\rangle$ gives the coupled block equations
\begin{equation}
H_{PP}|\psi_P\rangle + H_{PQ}|\psi_Q\rangle = E|\psi_P\rangle, \qquad
H_{QP}|\psi_P\rangle + H_{QQ}|\psi_Q\rangle = E|\psi_Q\rangle,
\label{eq:block_eqs}
\end{equation}
where $|\psi_P\rangle \equiv P|\psi\rangle$, $|\psi_Q\rangle \equiv Q|\psi\rangle$, and $H_{XY}\equiv XHY$ for $X,Y\in\{P,Q\}$.

\noindent Therefore,
\[
H_{PP}|\psi_P\rangle + H_{PQ}|\psi_Q\rangle = E|\psi_P\rangle,\quad
H_{QP}|\psi_P\rangle + H_{QQ}|\psi_Q\rangle = E|\psi_Q\rangle.
\]
Solving the second equation,
\[
|\psi_Q\rangle = \frac{1}{E-H_{QQ}}H_{QP}|\psi_P\rangle,
\]
and substituting back gives
\[
\left(H_{PP} + H_{PQ}\frac{1}{E-H_{QQ}}H_{QP}\right)|\psi_P\rangle = E|\psi_P\rangle,
\]
with the effective Hamiltonian
\[
H_{\text{eff}} = H_{PP} + H_{PQ}\frac{1}{E-H_{QQ}}H_{QP}.
\]

\noindent
We are interested in the low energy sector $E \approx E_0$ and $E \approx E_1$,
so we take $E = \frac{E_0+E_1}{2}$.

\begin{align}
H_{\text{eff}} \approx H_{PP} + H_{PQ}\frac{1}{\frac{E_0+E_1}{2}-H_{QQ}}H_{QP}
\end{align}

\noindent where $E_0, E_1$ are the dominant diagonal elements of the low energy $(2\times2)$ sector. To avoid inversion of $(E-H_{QQ})$, it is better to have a diagonal $H_{QQ}$ with $\text{diag}(E_3, E_4)$ matrix elements.

This is achieved by rotating the $T_-$ and $T_+$ states to obtain a $T_{\text{low}} = T_l$ and $T_{\text{high}} = T_h$ basis. So that the low energy qubit subspace will consist of $S_0$ and $T_l$ and the other states $(T_h, T_0)$ form the higher energy sector.

Therefore, to simplify the calculation further, we rotate and diagonalize the $T_+, T_-$ block to define:

\begin{align}
T_L &= c_-^L |T_-\rangle + c_+^L |T_+\rangle \\
\text{and } T_h &= c_-^h |T_-\rangle + c_+^h |T_+\rangle
\end{align}

\noindent with
\begin{align}
c_-^L &= \frac{a+\lambda}{\sqrt{b^2+(a+\lambda)^2}} & c_+^L &= \frac{-b}{\sqrt{b^2+(a+\lambda)^2}} \\
c_-^h &= \frac{a-\lambda}{\sqrt{b^2+(a-\lambda)^2}} & c_+^h &= \frac{-b}{\sqrt{b^2+(a-\lambda)^2}}
\end{align}

\noindent where 
\begin{align}
a &= \mu_B \bar{g} B_z \\
b &= \gamma^2/4 \\
\lambda &= \sqrt{a^2+b^2} = \sqrt{(\mu_B\bar{g} B_z)^2+(\gamma^2/4)^2}
\end{align}

\noindent
In the basis $\{S_0, T_l, T_h, T_0\}$, the Hamiltonian becomes:

% \begin{widetext}
\begin{table}[h]
\centering
\caption{Effective Hamiltonian matrix in the rotated basis.}
\label{tab_tptm_Ham}
\begin{tabular}{c|cccc}
 & $S_0$ & $T_l$ & $T_h$ & $T_0$ \\
\hline
$S_0$ & $-\frac{3J}{4} - \frac{\gamma^2}{4}$ & $(c_+^L-c_-^L)i\frac{D\sqrt{2}}{4}$ & $(c_-^h-c_+^h)i\frac{D\sqrt{2}}{4}$ & $\mu_B\frac{\Delta g}{2}$ \\
$T_L$ & $(c_+^L-c_-^L)i\frac{D\sqrt{2}}{4}$ & $\frac{J}{4}-\sqrt{(\frac{\gamma^2}{4})^2+(\mu_B B_z\bar{g})^2}$ & 0 & 0 \\
$T_h$ & $(c_-^h-c_+^h)i\frac{D\sqrt{2}}{4}$ & 0 & $\frac{J}{4}+\sqrt{(\frac{\gamma^2}{4})^2+(\mu_B B_z \bar{g})^2}$ & 0 \\
$T_0$ & $\mu_B\frac{\Delta g}{2}$ & 0 & 0 & $\frac{J}{4}+\frac{\gamma^2}{4}$ \\
\end{tabular}
\end{table}
% \end{widetext}

In terms of the projectors to the low energy subspace defined earlier, $H$ assumes the form,

\begin{table}[H]
\centering
\begin{tabular}{c|cccc}
 & $S_0$ & $T_l$ & $T_h$ & $T_0$ \\
\hline
$S_0, T_l$ & \multicolumn{2}{c|}{{$PHP$}} & {{$PHQ$}} \\
 & \multicolumn{2}{c|}{} & \multicolumn{2}{c}{} \\
\hline
$T_h, T_0$ & \multicolumn{2}{c|}{{$QHP$}} & {{$QHQ$}} \\
 & \multicolumn{2}{c|}{} & \multicolumn{2}{c}{} \\
\end{tabular}
\end{table}

Generally then,
\begin{align}
H &= \begin{pmatrix}
E_1 & \delta_{12} & \delta_{13} & \delta_{14} \\
\delta_{12}^* & E_2 & \delta_{23} & \delta_{24} \\
\delta_{13}^* & \delta_{23}^* & E_3 & \delta_{35} \\
\delta_{14}^* & \delta_{24}^* & \delta_{34}^* & E_4
\end{pmatrix}
\end{align}

\noindent In the basis formed by the projectors
\[
PHP=\begin{pmatrix} E_1 & \delta_{12} \\ \delta_{12}^* & E_2 \end{pmatrix},\;
PHQ=\begin{pmatrix} \delta_{13} & \delta_{14} \\ \delta_{23} & \delta_{24} \end{pmatrix},\;
QHP=\begin{pmatrix} \delta_{13}^* & \delta_{23}^* \\ \delta_{14}^* & \delta_{24}^* \end{pmatrix},\;
QHQ=\begin{pmatrix} E_3 & 0 \\ 0 & E_4 \end{pmatrix}\, ,
\]

\noindent where
\begin{align}
E_1 &= -\frac{3J}{4} - \frac{\gamma^2}{4} \\
E_2 &= \frac{J}{4} - \sqrt{(\frac{\gamma^2}{4})^2 + (\mu_B B_z \bar{g})^2} \\
E_3 &= \frac{J}{4} + \sqrt{(\frac{\gamma^2}{4})^2 + (\mu_B B_z \bar{g})^2} \\
E_4 &= \frac{J}{4} + \frac{\gamma^2}{4}
\end{align}

\noindent and
\[
\delta_{13}=(c_-^h-c_+^h)i\frac{D\Omega}{4},\;
|\delta_{13}|^2=(c_-^h-c_+^h)^2\frac{D^2}{8},\;
\delta_{14}=\mu_B\frac{\Delta g}{2},\;
|\delta_{14}|^2=\mu^2 B^2\frac{(\Delta g)^2}{4},\;
\delta_{23}=0,\;
\delta_{24}=0
\]

\noindent
Thus, the effective $2 \times 2$ Hamiltonian becomes:

\begin{eqnarray}
H_{\text{eff}}^{2\times2} &=& A_1 \mathbf{1} + A_3 \sigma_z + A_2 \sigma_y \label{Eq_qubit_ham} \\
    A_1 &=& \frac{1}{8} \Biggl\{ -\frac{2 D^2 \left(-\sqrt{16 \mu_B^2 B_z^2 \bar{g}^2+\gamma^4}+4\mu_B B_z \bar{g}  +\gamma^2\right)^2}{\left(4\mu_B B_z \bar{g}  \left(4\mu_B B_z \bar{g} -\sqrt{16\mu_B^2 B_z^2 \bar{g}^2+\gamma^4}\right)+\gamma^4\right) \left(3 \sqrt{16\mu_B^2 B_z^2 \bar{g}^2+\gamma^4}+\gamma^2+4 J\right)}\nonumber \\ 
    &~& -\sqrt{16\mu_B^2 B_z^2 \bar{g}^2+\gamma^4} -\frac{8 B^2 \text{$\Delta $g}^2 \mu_B
   ^2}{\sqrt{16\mu_B^2 B_z^2 \bar{g}^2+\gamma^4}+3 \gamma^2+4 J}-\gamma^2-2 J\Biggr\}\\
    A_3 &=& \frac{1}{8} \Biggl\{ -\frac{2 D^2 \left(-\sqrt{16 \mu_B^2 B_z^2 \bar{g}^2+\gamma^4}+4\mu_B B_z \bar{g}  +\gamma^2\right)^2}{\left(4\mu_B B_z \bar{g}  \left(4\mu_B B_z \bar{g} -\sqrt{16\mu_B^2 B_z^2 \bar{g}^2+\gamma^4}\right)+\gamma^4\right) \left(3 \sqrt{16\mu_B^2 B_z^2 \bar{g}^2+\gamma^4}+\gamma^2+4 J\right)}\nonumber \\ 
    &~& +\sqrt{16\mu_B^2 B_z^2 \bar{g}^2+\gamma^4} -\frac{8 B^2 \text{$\Delta $g}^2 \mu_B
   ^2}{\sqrt{16\mu_B^2 B_z^2 \bar{g}^2+\gamma^4}+3 \gamma^2+4 J}-\gamma^2-4 J\Biggr\}\\
   A_2 &=& -\frac{D \left(\sqrt{16\mu_B^2 B_z^2 \bar{g}^2+\gamma^4}+4 \mu_B B_Z \bar{g}  +\gamma^2\right)}{4 \sqrt{4 \mu_B B_Z \bar{g} \left(\sqrt{16\mu_B^2 B_z^2 \bar{g}^2+\gamma^4}+4 \mu_B B_Z \bar{g}\right)+\gamma^4}}
\end{eqnarray}

\noindent where the qubit parameters in terms of the bare fermionic parameters are
\begin{align}
J &= \frac{4U}{U^2-\varepsilon^2}(t_\perp^2 - t_{\perp SO}^2) \\
D &= \frac{8U}{U^2-\varepsilon^2}(t_\perp \times t_{\perp SO}) \\
\gamma_\perp^2 &= \frac{8U}{U^2-\varepsilon^2}(t_{\perp SO})^2
\end{align}

\noindent
This completes the derivation of the effective qubit Hamiltonian for a ladder rung under SOC and magnetic field. The energy gap can then be obtained by diagonalising the Hamiltonian and taking the difference of the eigenvalues, which gives
\begin{equation}
    \Delta E = 2\sqrt{A_3^2 + A_2^2} .
    \label{eq:gap_vs_barrier}
\end{equation}
This is the expression which we fit to experimental data of the energy gap as a function of barrier gate voltage (Fig.~\ref{fig:Jperp}), in order to extract the relevant theoretical parameters. To be more precise, we use the independently calibrated local $g$-factors, set the on-site detuning $\epsilon=0$ (as in the experiment), and fit $t_\perp$, $t_{\perp SO}$ and $U$, assuming an exponential relation between $t_\perp$ and barrier voltage (Fig.~\ref{fig:Jperp}). The data is well fitted by the functional form of Eq.~\ref{eq:gap_vs_barrier}.

Here we bring the table of the extracted parameters
\begin{table}[h]
\centering
\caption{Fitted model parameters for three rungs and four rungs.}
\label{tab:fitted_params}
\begin{tabular}{lcc}
\toprule
Parameter & Three rungs & Four rungs \\
\midrule
$U$ (GHz)        & 333.6722 & 300.0000 \\
$g_1$            & 0.3700   & 0.3700   \\
$g_2$            & 0.3700   & 0.3700   \\
$t_{\perp}^{SO}$ (GHz)   & 0.3167   & 0.3000   \\
$A$ (GHz)        & 1.0000   & 1.1445   \\
$\alpha$ (1/mV)  & 0.00504  & 0.02234  \\
$V_0$ (mV)       & 77.6775  & 5.1983   \\
$t_{\infty}$ (GHz) & 0.0000 & 0.0000   \\
$\epsilon$ (GHz) & 0.0000   & 0.0000   \\
\bottomrule
\end{tabular}
\end{table}

\subsubsection{Leg Parameter Estimation}

We consider a two-spin Hamiltonian describing a single leg bond (leg index suppressed for brevity),
\begin{equation}
H =
J_{\parallel}\,\mathbf S_0 \cdot \mathbf S_1
+
D^z_{\parallel}\left(S_{y0}S_{x1} - S_{x0}S_{y1}\right)
+
(\gamma^z)^2 S^z_0 S^z_1
+
\mu_B B_z \left(g_0 S^z_0 + g_1 S^z_1\right),
\end{equation}
where the microscopic parameters $J_{\parallel}, D_{\parallel}^z$, and $(\gamma_{\parallel}^z)^2$ are functions of the unknown leg tunneling amplitudes, while $U$ and $\epsilon$ are known from the previous subsection on rung parameter estimations. For the numerical fits, we keep $D_{\parallel}^z$ and $(\gamma_{\parallel}^z)^2$, although spin-orbit coupling along the legs is very weak, as stated elsewhere. The fact that spin-orbit coupling is small is consistent with the fact that the parameters extracted from the fits barely change when $D_{\parallel}^z$ and $(\gamma_{\parallel}^z)^2$ are neglected when fitting the data.

Projecting onto the singlet--triplet basis $\{|S\rangle,|T_0\rangle,|T_+\rangle,|T_-\rangle\}$, the $|S\rangle$ and $|T_0\rangle$ states form a closed subspace, while $|T_\pm\rangle$ remain decoupled. In the $\{|S\rangle,|T_0\rangle\}$ sector the Hamiltonian reads
\begin{equation}
H_{S/T_0}
=
\begin{pmatrix}
-\frac{3}{4}J_{\parallel} - \frac{(\gamma_{\parallel}^z)^2}{4}
&
\frac{\mu_B B_z (g_0-g_1)}{2} + \frac{iD_{\parallel}^z}{2}
\\[6pt]
\frac{\mu_B B_z (g_0-g_1)}{2} - \frac{iD_{\parallel}^z}{2}
&
\frac{1}{4}J_{\parallel} - \frac{(\gamma_{\parallel}^z)^2}{4}
\end{pmatrix}.
\end{equation}

Equivalently, this can be written in Pauli-matrix form as
\begin{equation}
H_{S/T_0}
=
-\frac{J_{\parallel}}{2}\sigma^z
+
\frac{D_{\parallel}^z}{2}\sigma^y
+
\frac{\mu_B B_z (g_0-g_1)}{2}\sigma^x
+
\left(-\frac{J_{\parallel}}{4} - \frac{(\gamma_{\parallel}^z)^2}{4}\right)\mathbb{I}.
\end{equation}

The resulting singlet--triplet splitting measured in spectroscopy is
\begin{equation}
\Delta E
=
\sqrt{
J_{\parallel}^2
+
\left(D_{\parallel}^z\right)^2
+
\mu_B^2 B_z^2 (g_0-g_1)^2 }
\end{equation}

\vspace{2mm}

We recall that the microscopic origin of the effective couplings is set by tunneling processes on each leg,
\begin{align}
J_{\parallel} &= \frac{4U}{U^2-\epsilon^2}\left(t_{\parallel}^2 - t_{\parallel,\mathrm{SO}}^2\right), \\
D_{\parallel}^z &= \frac{8U}{U^2-\epsilon^2}\, t_{\parallel} t_{\parallel,\mathrm{SO}}, \\
(\gamma_{\parallel}^z)^2 &= \frac{8U}{U^2-\epsilon^2}\, t_{\parallel,\mathrm{SO}}^2,
\end{align}
where $t_{\parallel}$ and $t_{\parallel,\mathrm{SO}}$ denote spin-conserving and spin--orbit-assisted tunneling amplitudes, respectively. The values of $t_{\parallel}$, $t_{\parallel,\mathrm{SO}}$, and the Zeeman asymmetry $g_0-g_1$ will be extracted now from experiment.

% The barrier voltage controls the spin-conserving tunneling via a monotonic dependence
% \begin{equation}
% t_{\parallel}(V) = A\,e^{-\alpha(V - V_0)} + t_{\parallel,\infty},
% \end{equation}
% providing a direct mapping between gate voltage and exchange strength.

We once again assume an exponential dependence of the spin-conserving tunneling on barrier gate voltage:
\begin{equation}
t(V) = A\,e^{-\alpha(V - V_0)} + t_{\infty} \,.
\end{equation}

The parameter extraction proceeds as follows. In the high-field regime,
$\mu_B |B_z(g_0-g_1)| \gg \{J_{\parallel},D_{\parallel}^z\}$, the dispersion reduces to
\begin{equation}
\hbar \omega(B) \approx \mu_B |g_0-g_1|\,|B_z|,
\end{equation}
allowing a direct determination of the Zeeman gradient $g_0-g_1$, performed in a regime where exchange is strongly suppressed by high barrier voltage.

At low magnetic field,
\begin{equation}
\hbar \omega(0) = \sqrt{J_{\parallel}^2 + (D_{\parallel}^z)^2},
\end{equation}
which is evaluated as a function of barrier voltage. The exchange $J_{\parallel}(V)$ is obtained from fits to the full magnetic-field-dependent dispersion at each voltage, while $D_{\parallel}^z$ follows from the residual splitting at zero field.

Finally, the extracted $J_{\parallel}(V)$ is fitted to the tunneling model above, yielding $t_{\parallel}(V)$. The (small) spin--orbit tunneling amplitude $t_{\parallel,\mathrm{SO}}$ is treated as a voltage-independent parameter.

\subsubsection{Numerical Calculations}

For all plots in Fig.~2, the theoretical predictions are obtained by directly solving the Hamiltonian in Eq.~\eqref{Eq_spin_ham} using exact diagonalization as implemented in the QuSpin package~\cite{weinberg2017quspin,weinberg2019quspin2}, with the experimentally estimated parameters as described above. We work in the rung singlet--triplet basis, constructed from the local two-spin eigenstates on each rung. The Hamiltonian is then diagonalized within this basis. We have also verified that, in the limit of large $U$ and at half filling, the low-energy spectrum of the Fermi--Hubbard ladder Hamiltonian in Eq.~\eqref{eq:fermion_ham_tb} coincides with that of the spin-only Hamiltonian in Eq.~\eqref{Eq_spin_ham}, justifying its sufficiency. Finally, the effective temperature $T$ of the ground-state manifold is adjusted to achieve improved agreement with the experimental data. We also note that we use the average spin gaps in Figs.~2 to~4 in the main, considering that the rung-to-rung variations are small (see Fig.~\ref{fig:Jperp}).

Regarding comparison with experimental data, the theoretical curves do not include SPAM errors. Although we mitigate readout errors using MLE method, we do not correct for initialization errors. Moreover, the readout-error matrix indicates that states with a larger number of triplets are more difficult to read out (see~\ref{sec:virtual}). 

Finally, during the experimental runs, large-amplitude voltage pulses are applied to an extensive set of gates. Therefore the imperfect virtual-gate calibration can cause the deviation from the expected curves ~\cite{Jirovec2025MitigationExchangeCrosstalk}.

\clearpage

\subsection{Bose--Hubbard Mapping, Transverse Correlator, and The Canted Antiferromagnet}
\label{sec:BEC}
In this section, our goal is to understand the spontaneous symmetry-breaking physics from the perspective of the equivalent bosonic model. To this end, we first consider the SOC-free limit, $D_\perp^x=0$. The effects of SOC have previously been discussed and will be reintroduced later. In this limit, Eq.~\eqref{eqn:effspinchainham} reduces to
\begin{equation}
H =
J_{\parallel}\sum_i \left(\widetilde S_i^x \widetilde S_{i+1}^x + \widetilde S_i^y \widetilde S_{i+1}^y\right)
+\frac{J_{\parallel}}{2}\sum_i \widetilde S_i^z \widetilde S_{i+1}^z
+\sum_i \widetilde h_{z,i}\,\widetilde S_i^z,
\label{eq:xxz}
\end{equation}
with $\widetilde h_{z,i}=h_{z,i}-J_\perp-J_\parallel/2$.

{A convenient interpretation of Eq.~\eqref{eq:xxz} is obtained through the mapping \cite{Matsubara1956,Giamarchi2003},
\begin{equation}
\widetilde S_i^+ = b_i^\dagger,
\qquad
\widetilde S_i^- = b_i,
\qquad
\widetilde S_i^z = n_i-\frac12,
\qquad
n_i=b_i^\dagger b_i\in\{0,1\},
\label{eqn:spbosemap}
\end{equation}
which identifies the effective spin degrees of freedom with hard-core bosons
(triplons). We note that Eq.~\eqref{eqn:spbosemap} should be understood as a mapping to a hard-core Bose--Hubbard model, rather than the conventional soft-core Bose--Hubbard model. The hard-core constraint is inherited directly from the original spin-(1/2) degrees of freedom: the two spin states $|\downarrow\rangle$ and $|\uparrow\rangle$ map onto the bosonic occupation states $n_i=0$ and $n_i=1$, respectively. Therefore, double occupation is forbidden, $n_i\in{0,1}$, and the local Hilbert space remains two-dimensional. Equivalently, this constraint can be viewed as arising from an infinite on-site repulsion $U\rightarrow\infty$ that excludes multiple boson occupancy.} Substituting these relations into Eq.~\eqref{eq:xxz} and ignoring constant terms yields,
\begin{equation}
H=
-t\sum_i\left(b_i^\dagger b_{i+1}+\mathrm{h.c.}\right)
+V\sum_i n_i n_{i+1}
-\sum_i \mu_i n_i,
\label{eq:bh}
\end{equation}
with
\begin{equation}
t=-\frac{J_\parallel}{2},
\qquad
V=\frac{J_\parallel}{2},
\qquad
\mu_i=-\widetilde h_{z,i}+V.
\label{eqn:bhm}
\end{equation}

It is easy to note that the $-\frac{D_\perp^x}{4i\sqrt2}\sum_i\left(b_i^\dagger-b_i\right)$ originates from the SOC-induced DM interaction $-\frac{D^x_\perp}{2\sqrt{2}}
\sum_{i}^{L}
\tilde{S}_i^y$, that we have momentarily ignored in Eq.~\eqref{eqn:effspinchainham}. In the bosonic representation it explicitly breaks the global $U(1)$ symmetry by creating or annihilating a boson, and therefore does not conserve particle number.

In the bosonic representation of Eq.~\eqref{eqn:bhm}, the applied magnetic field enters as an effective chemical potential for triplons, whereas the transverse exchange interaction generates nearest-neighbor hopping of the hard-core bosons. Since $h_z$ and $J_{\parallel}$ are held fixed, varying $J_{\perp}$ effectively tunes the chemical potential $\mu$. The two quantum critical points of the ladder therefore map onto critical chemical potentials $\mu_{c1}$ and $\mu_{c2}$. In the intermediate regime, $\mu_{c1}<\mu<\mu_{c2}$, the bosonic system is gapless and
compressible, corresponding to the canted antiferromagnetic (CAFM) phase in the spin
representation. 

The Bose--Hubbard mapping allows this phase to be viewed as a 1D analogue of a condensate
of hard-core bosons, providing a direct connection between magnetic order and bosonic
coherence. As this is a one-dimensional system, the intermediate phase exhibits only
quasi-long-range order. We therefore characterize it through the transverse and longitudinal
correlators of the associated Luttinger liquid \cite{CoupledLaddersMagneticField},
\begin{align}
\langle \widetilde S^+(x,\tau)\,\widetilde S^-(0,0)\rangle
&\sim \cos(2\pi \tilde m x)\left(\frac{1}{r}\right)^{2K+1/(2K)}
+ \cos(\pi x)\left(\frac{1}{r}\right)^{1/(2K)} \sim \cos(\pi x)\left(\frac{1}{r}\right)^{1/(2K)},
\label{eq:transverse_corr}
\\
\langle \widetilde S^z(x,\tau)\,\widetilde S^z(0,0)\rangle
&\sim \tilde m^2 + \frac{1}{r^2}
+ \cos\!\big(\pi(1-2\tilde m)x\big)\left(\frac{1}{r}\right)^{2K},
\label{eq:longitudinal_corr}
\end{align}
where $r=\sqrt{x^2+(u\tau)^2}$ (reducing to $r=|x|$ at equal time and x is distance between sites, $u$ is the spin-wave
(sound) velocity of the effective Luttinger liquid, $K \leq 1$ is the Luttinger parameter
controlling the anomalous decay exponents, and $\tilde m = \langle \widetilde S^z\rangle$ is
the magnetization of the effective spin-1/2 chain, set by the applied field \cite{CoupledLaddersMagneticField}. Within
Eq.~\eqref{eq:transverse_corr}, the incommensurate-envelope term always decays faster than
the $\pi$-staggered term, 
the staggered piece $\cos(\pi x)\,r^{-1/(2K)}$ therefore controls the long-distance
behavior and its staggered form describes alternating transverse moments on neighboring
sites with algebraically decaying phase coherence: this is what makes the phase ``canted''
(transverse spin components) and ``antiferromagnetic'' (staggered at wavevector $\pi$).
Comparing this exponent to the oscillating piece of the longitudinal correlator
Eq.~\eqref{eq:longitudinal_corr}, $\cos\!\big(\pi(1-2\tilde m)x\big)\,r^{-2K}$: since
$1/(2K) < 2K$ whenever $K>1/2$, and $K$ remains in the range $(3/4,1)$ throughout the
gapless phase \cite{CoupledLaddersMagneticField} --- always above this crossover value ---
the transverse exponent $1/(2K)$ is smaller than the longitudinal exponent $2K$, so the
transverse (phase-coherence) correlator decays more slowly and dominates over the
longitudinal (density-wave) one.

From the bosonic perspective, this same algebraically decaying transverse correlator
corresponds to a quasi-condensate of hard-core bosons, with coherence captured by the
one-body density matrix
\begin{equation}
\rho^{(1)}_{ij} = \langle b_i^\dagger b_j\rangle,
\end{equation}
which measures phase coherence between different sites. Using the
spin to hard-core boson mapping,
\begin{equation}
\langle b_i^\dagger b_j\rangle = \langle \widetilde S_i^+\widetilde S_j^-\rangle,
\label{eq:coherence}
\end{equation}
showing that condensate coherence is encoded directly in the transverse spin
correlator. By contrast,
\begin{equation}
\langle \widetilde S_i^z\widetilde S_j^z\rangle
=
\Big\langle
\left(n_i-\frac12\right)
\left(n_j-\frac12\right)
\Big\rangle,
\end{equation}
is a density--density correlator [Eq.~\eqref{eq:longitudinal_corr}] and probes longitudinal
spin-density modulations rather than phase coherence.
This distinction is important when characterizing the intermediate phase.
Longitudinal correlations identify the incommensurate spin-density-wave
structure of the magnetization, whereas transverse correlations probe the
coherence of the bosonic degrees of freedom. The CAFM phase is therefore most
naturally associated with enhanced transverse correlations, corresponding to a
quasi-condensed hard-core boson state in the equivalent Bose--Hubbard picture.
To quantify this coherence we construct the transverse correlation matrix numerically from Eqn.~\eqref{Eq:mainH1}
\begin{equation}
M_{ij} = \langle \widetilde S_i^+\widetilde S_j^-\rangle,
\qquad i\neq j,
\label{eqn:transcorr}
\end{equation}
and use its largest eigenvalue as an order-parameter proxy. 
A large leading eigenvalue indicates that the correlations are
dominated by a single coherent mode and therefore provides a direct measure of
condensate coherence.

The diagonal elements are omitted because
\begin{equation}
M_{ii} = \langle \widetilde S_i^+\widetilde S_i^-\rangle
\end{equation}
contain only local density information and do not probe coherence between
different sites. Restricting the analysis to the off-diagonal sector therefore
isolates the phase-coherent part of the correlation matrix.

\begin{figure}[h]
\centering
\includegraphics[width=0.3\textwidth]{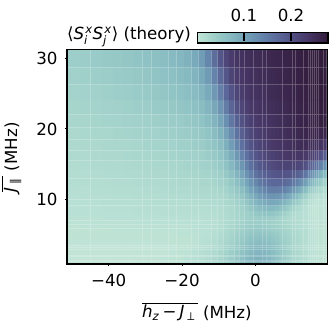}
\caption{%
\textbf{Transverse correlator plot, showing the region with supposed CAFM}
As follows from theory, the CAFM phase is characterized by transverse AFM correlations as defined in Eq.~\eqref{eqn:transcorr}. Here, the theoretical plot shows that the region with this type of correlation essentially reproduces the structure of the phase diagram. Although in the main text we refer to the intermediate phase as CAFM-like, the longitudinal correlations detect only the spin-density-wave mode of the intermediate regime. The theoretical plot here confirms that, with additional rotations in the xy-plane, we would also observe phase coherence.
}
\label{fig:BEC}
\end{figure}

In Fig.~\ref{fig:BEC} in presence of SOI (restored), we observe that the region identified as
CAFM-like in the phase diagram coincides with the regime where the leading
eigenvalue of $M$ is enhanced, consistent with the expected emergence of strong
transverse coherence in the equivalent hard-core boson description.

\clearpage

\subsection{Slope extraction from phase diagram}
\label{sec:slopes}
We extract the slopes of the boundaries seen in the experimentally measured phase diagram, so we can compare them with the theoretically expected slopes (Section~\ref{sec:theory}).
We determine the slope of the RS-to-FP boundary over a range of $J_\parallel$ of $[10, 31]$ MHz, and for the CAFM-to-FP boundary the range is $[9, 17]$ MHz.

The phase boundary is identified by fitting a sigmoid function to the triplet number profile at each value of $J_\parallel$, using the plateau values deep in each phase, $M_\mathrm{RS}$, $M_\mathrm{CAFM}$, $M_\mathrm{FP}$. The left boundary is fitted with:
\begin{equation}
    f(x) = M_\mathrm{RS} + \frac{M_\mathrm{CAFM} - M_\mathrm{RS}}
    {1 + e^{-(x - x_0)/W}},
    \label{eq:sigmoid_left}
\end{equation}
where $x_0$ is the inflection point -- the position where the transition is steepest -- and $W$ is a measure of the transition width. Both are free parameters of the fit. 

\begin{figure}[H]
\centering
\includegraphics[width=\textwidth]{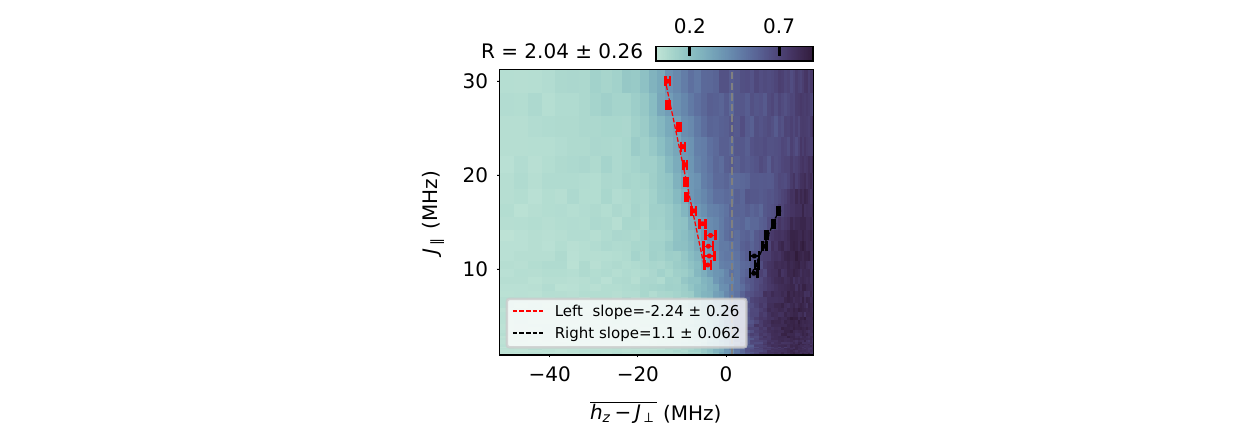}
\caption{%
\textbf{Slope extraction from the phase diagram.}
The phase diagram is taken directly from the main body of the paper and shows the experimentally measured average triplet number as a function of $\overline{h_z - J_{\perp}}$ and $\overline{J_{\parallel}}$. Scatter points are extracted by fitting a sigmoid to the slope and identifying the sharpest inflection point, corresponding to the phase boundary. Red and black dashed lines are weighted linear fits that provide the slope values.
}
\label{fig:slopes}
\end{figure}

Similarly, the right boundary is fitted with:
\begin{equation}
    f(x) = M_\mathrm{CAFM} + \frac{M_\mathrm{FP} - M_\mathrm{CAFM}}
    {1 + e^{-(x - x_0)/W}}.
    \label{eq:sigmoid_right}
\end{equation}
In Fig.~\ref{fig:slopes} the scatter points show the extracted $x_0$ values, with error bars representing the $1\sigma$ uncertainty from the fit covariance.Then we perform a weighted linear fit through the extracted boundary points, where each point is weighted by the inverse of its fit uncertainty.

As is clear from the figure, the absolute value of the slope ratio $R$ is 
$2.04 \pm 0.26$, in agreement with the theoretical prediction of Sec.~\ref{sec:theory}.

\clearpage

\subsection{Additional phase diagram data}

\label{sec:PD_from_corr}

\begin{figure}[h]
\centering
\includegraphics[width=\textwidth]{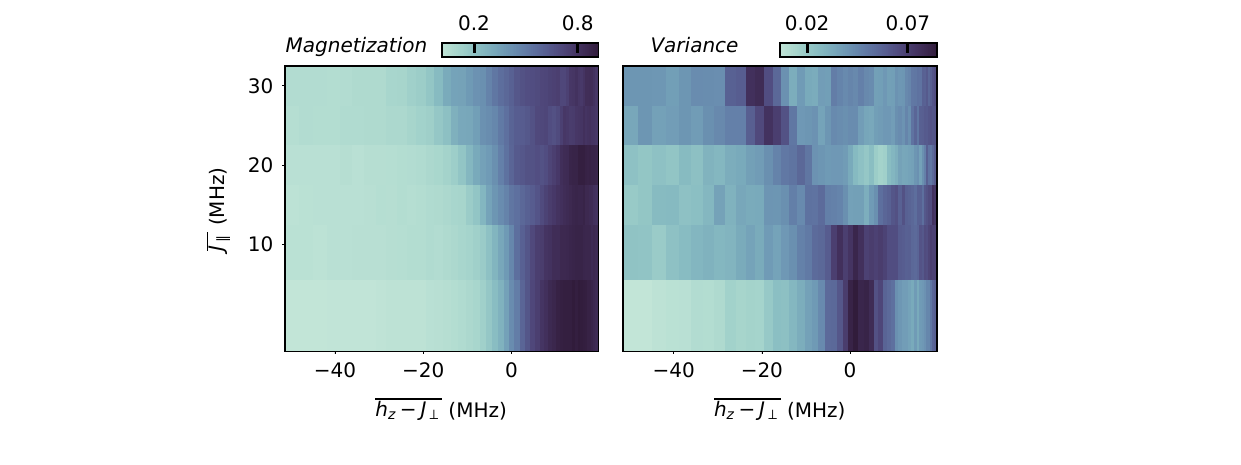}
\caption{%
\textbf{Phase diagram reconstructed from the data used to produce the connected correlators shown in Fig.~\ref{fig:fig4}.}
(a) triplet number as a function of $\overline{h_z - J_{\perp}}$ and $\overline{J_{\parallel}}$. We provide this dataset here because, for the measurements used to produce the correlation functions, individual SPAM matrices were obtained for each probed $\overline{J_{\parallel}}$. As a result, MLE provides significantly improved readout-error mitigation. However, only six values of $\overline{J_{\parallel}}$ were sampled, since this measurement is more time-consuming. With this improved readout-error mitigation, additional fine structure in the CAFM phase is resolved, presumably corresponding to first-order transitions between different triplet number plateaus that are also present in the theoretical plots shown in the main text.
(b) triplet number variance obtained from the same dataset. Here, the variance peaks are significantly sharper than for similar results in Fig. ~\ref{fig:fig3}. Altogether, this indicates that in our experiment the resolution of the variance and the additional steps in triplet number are mainly limited by readout errors rather than by strong SOI.
}
\label{fig:pd_correlation_data}
\end{figure}

\end{document}